\documentclass[11pt,a4paper]{article}
\usepackage{amsmath,dsfont}
\usepackage[normalem]{ulem}
\usepackage{jcappub}
\usepackage{hyperref}
\hypersetup{colorlinks=True,citecolor=blue}
\usepackage{amsfonts}
\usepackage{amssymb}
\usepackage{caption}
\usepackage{float}
\usepackage{bm}
\usepackage{graphicx}
\usepackage{color}
\usepackage{verbatim}

\usepackage{subfig}
\newcommand{\ber}{\begin{eqnarray}}
\newcommand{\eer}{\end{eqnarray}}

\def\ie{i.e.}

\def\im{{\rm i}}

\def\vphi{\varphi}

\def\mn{{Mon.\@ Not.\@ Roy.\@ Ast.\@ Soc.\ }}

\def\prl{{Phys.\@ Rev.\@ Lett.\ }}
\def\prd{{Phys.\@ Rev.\@ D\ }}

\def\plb {{Phys.\@ Lett.\@ B\ }}

\title{Sourcing Dark Matter and Dark Energy from $\alpha$-attractors}

\author[a]{Swagat S. Mishra,}

\author[a]{Varun Sahni}

\author[b,c]{and Yuri Shtanov}
\affiliation[a]{Inter-University Centre for Astronomy and Astrophysics, Post Bag 4, Ganeshkhind,  Pune 411 007, India}%
\affiliation[b]{Bogolyubov Institute for Theoretical Physics, Kiev 03680, Ukraine} %
\affiliation[c]{Department of Physics, Taras Shevchenko Kiev National University, Kiev, Ukraine} %
\emailAdd{swagat@iucaa.in} \emailAdd{varun@iucaa.in} \emailAdd{shtanov@bitp.kiev.ua}
\abstract{
In \cite{linde1}, Kallosh and Linde drew attention to a new family of superconformal
inflationary potentials, subsequently called $\alpha$-attractors \cite{linde2}. The
$\alpha$-attractor family can interpolate between a large class of inflationary models.
It also has an important theoretical underpinning within the framework of supergravity.
We demonstrate that the $\alpha$-attractors have an even wider appeal since they may
describe dark matter and perhaps even dark energy.
The dark matter associated with the $\alpha$-attractors, which we call $\alpha$-dark matter ($\alpha$DM),
 shares many of the attractive features
of {\em fuzzy\/} dark matter, with $V(\vphi) = \frac{1}{2}m^2\vphi^2$, while having none
of its drawbacks. Like fuzzy dark matter, $\alpha$DM can have a large Jeans length which
could resolve the cusp--core and substructure problems faced by standard cold dark
matter. $\alpha$DM  also has an appealing tracker property which enables it to converge
to the late-time dark matter asymptote, $\langle w\rangle \simeq 0$, from a wide range of
initial conditions. It thus avoids the enormous fine-tuning problems faced by the
$m^2\vphi^2$ potential in describing dark matter.}


\begin{document}
\maketitle

%\tableofcontents

\section{Introduction}\label{sec: intro}

Dark matter (DM) and dark energy (DE) are two of the most enigmatic observables in
current cosmology. While dark matter is assumed to be pressureless, dark energy is
believed to have large negative pressure which accelerates the present universe. Nothing
significantly more is known about the nature of either dark matter or dark
energy. It is accepted that dark matter must cluster, in order to account for
the missing-mass problem associated with individual galaxies and with galaxy clusters.
Furthermore, there are strong reasons to believe that dark matter must have already been
in place by redshifts of $z \gtrsim 10^{4}$. This is indicated by primordial fluctuations
in the cosmic microwave background (CMB), whose small amplitude ($\Delta T/T \sim
10^{-5}$) is suggestive of an equally small amplitude of primordial density fluctuations
at $z \gtrsim 10^3$. Such small fluctuations would have had difficulty in growing to the
much larger values today, $\delta\rho/\rho > {\cal O} (1)$, in a universe consisting
solely of baryons \cite{sahnicoles_95,dodelson}.

The origin of dark energy could, on the other hand, be much more recent, since there is
observational evidence to suggest that the universe commenced accelerating at $z \lesssim
{\rm few}$.

Theoretical models of dark matter usually subscribe to the view that it is made up of
hitherto undetected elementary particles called WIMPs (weakly interacting massive
particles). However, despite several decades of systematic searches by elaborate
experiments, a firm consensus on the existence of WIMPs has eluded researchers. Our
understanding of dark energy faces an even greater dilemma since its basic properties,
such as its pressure and density, have only been indirectly deduced via cosmological
observations of the expansion history, $H(z)$, or the luminosity distance, $D_L(z)$.
Consequently, it has even been suggested that both DM \& DE may owe their origin to a
modification of the laws of gravity on large scales \cite{sahni04,DE}.

In this paper, we move past the dominant paradigm of particle-like WIMP dark matter and
examine the alternative possibility that DM could have the structure of a scalar field.
Ever since the advent of Inflation, scalar field models have  played an increasingly
prominent role in our understanding of the very early universe. Scalar-field
models have also been advocated to describe dark energy \cite{DE}. Recently, Kallosh and
Linde \cite{linde1} have discussed a new class of scalar-field models, called
$\alpha$-attractors \cite{linde2}, which have an attractive feature of describing an
entire family of inflationary models within a common theoretical setting. In this paper,
we show how the $\alpha$-attractors may have an even wider appeal, since they can
describe dark matter, and perhaps even dark energy.

Our paper is structured as follows: section \ref{sec:alphainf} discusses the
$\alpha$-attractor Lagrangian  in the context of conformal inflation, while section
\ref{sec:alphadm} discusses the $\alpha$-attractor Lagrangian in the context of dark matter
and dark energy. Section \ref{sec:dm} explores the role of $\alpha$-attractors as dark
matter and includes a comprehensive discussion of gravitational instability.
 Section \ref{sec:DE} discusses $\alpha$-attractors in the context of dark energy. A
summary of our results is presented in section \ref{sec:summary}, in which we also discuss
the possibility that the inflaton could
play the role of dark matter at late times, after preheating.

\section{Conformal inflation}\label{sec:alphainf}

In \cite{linde1}, Kallosh \& Linde drew attention to an interesting new family
of potentials which could give rise to successful inflation. %A generalization
%of this family was called $\alpha$-attractors in \cite{linde2}.
The starting point of this discovery was the observation that %the $SO(1,1)$
%symmetric
the action with the Lagrangian
\ber
{\cal L} = \sqrt{-g}\left[ \frac{1}{2}\partial_\mu\chi\partial^\mu\chi +
\frac{\chi^2}{12}R(g) - \frac{1}{2}\partial_\mu\phi\partial^\mu\phi
-\frac{\phi^2}{12}R(g) - \frac{\tilde\lambda }{4}\left (\phi^2 - \chi^2\right )^2\right]
\, , \label{eq:1}
\eer
where $\chi$ and $\phi$ are scalar fields, and $\tilde\lambda$ is a dimensionless
constant, is invariant under the ${\rm O}(1, 1)$ group of transformations in the $(\chi,
\phi)$ space and also under the group of local conformal transformations. When the local
conformal gauge is fixed to
\ber \label{eq:gauge-fix}
\chi^2-\phi^2 = 6 m_{p}^2 \, ,
\eer
this Lagrangian can be parameterized by
\ber
\chi = \sqrt{6} m_p \cosh{\frac{\vphi}{\sqrt{6} m_p}}~, \quad \phi = \sqrt{6} m_p
\sinh{\frac{\vphi}{\sqrt{6} m_p}} \, ,
\eer
and reduces to
\ber
L = \sqrt{-g}\left\lbrack \frac{m_{p}^2}{2} R -
\frac{1}{2}\partial_\mu\vphi\partial^\mu\vphi - 9\tilde\lambda  m_{p}^4 \right\rbrack \,
, \label{eq:deS}
\eer
%Eq. (\ref{eq:deS})
describing Einstein's gravity with the reduced Planck mass $m_p = 1/ \sqrt{8 \pi G}
\approx 2.4 \times 10^{18}$~GeV and cosmological constant $\Lambda = 9\tilde\lambda
m_{p}^2$. A conformally invariant generalization of (\ref{eq:1}) $\tilde\lambda  \to
F(\phi/\chi) /9$, with an arbitrary function $F$ that deforms the ${\rm O} (1, 1)$
symmetry, results in the standard Lagrangian
\ber
{\cal L} = \sqrt{-g}\left\lbrack \frac{m_{p}^2}{2} R -
\frac{1}{2}\partial_\mu\vphi\partial^\mu\vphi - V (\varphi) \right\rbrack
\eer
with a scalar-field potential
\ber \label{eq:pot}
V (\varphi) = m_{p}^4 F \left( \tanh{\frac{\vphi}{\sqrt{6} m_p}} \right) \, .
\eer
%which describes
In \cite{linde1}, various canonical potentials $V(\vphi)$ were investigated in the
context of inflation. In \cite{linde2}, a family of potentials called $\alpha$-attractors
were considered following the prescription $V (\varphi) \to V \left( \varphi /
\sqrt{\alpha} \right)$ starting from canonical potentials $V (\varphi)$ in (\ref{eq:pot}).
%\ber
%V(\vphi) = F\left (\tanh{\frac{\vphi}{\sqrt{6\alpha} M}}\right ). \label{eq:alpha}
%\eer
The $\alpha$-attractors are able to parameterize a wide variety of inflationary settings
including chaotic inflation and the Starobinsky model. (The parameter $\alpha$ was
shown to be related to the curvature of the superconformal K$\ddot{\rm a}$hler metric in
\cite{linde2}.)

\section{Conformal dark matter}\label{sec:alphadm}

In this paper, we would like to demonstrate that the %$\alpha$-attractor
theory of the previous section %with the family of potentials (\ref{eq:pot})
can also describe dark matter and perhaps even dark energy.  To this end, we first
incorporate the scalar-field Lagrangian
\ber
{\cal L} = \sqrt{-g}\left[ \frac{1}{2}\partial_\mu\chi\partial^\mu\chi +
\frac{\chi^2}{12}R(g) - \frac{1}{2}\partial_\mu\phi\partial^\mu\phi
-\frac{\phi^2}{12}R(g) - \frac{F (\phi/\chi)}{36 m_{p}^4}\left (\phi^2 - \chi^2\right
)^2\right] \label{eq:extended}
\eer
into the Standard Model (SM) of electroweak interactions respecting the local conformal
invariance.  Since the only sector in the SM that breaks conformal invariance is the
kinetic and potential terms of the Higgs field $h$, we modify the relevant parts in the
minimal possible way:
\ber \label{eq:lag-or}
{\cal L}_h = - \sqrt{- g} \left(\frac12 \partial_\mu h \partial^\mu h + \frac{h^2}{12} R
+ \frac{\lambda_h}{4} \left[ h^2 - \frac{\upsilon^2}{6 m_{p}^2} \left(\chi^2 - \phi^2 \right)
\right]^2 \right) \, ,
\eer
where $\upsilon \approx 246$~GeV is the Higgs-field vacuum expectation value, and
$\lambda_h$ is its self-coupling constant. After fixing the conformal gauge as in
(\ref{eq:gauge-fix}), we obtain a theory with the Lagrangian
\ber \label{eq:lag-eff}
{\cal L} = \sqrt{-g}\left[ \frac{m_{p}^2 - h^2/6}{2} R - \frac{1}{2} \partial_\mu \vphi
\partial^\mu \vphi - V (\varphi) %- \frac12 \partial_\mu h \partial^\mu h -
%\frac{\lambda_h}{4} \left( h^2 - \upsilon^2\right)^2
\right]  + {\cal L}_{\rm SM} \, ,
\eer
where $V (\varphi)$ is given by (\ref{eq:pot}), and ${\cal L}_{\rm SM}$ is the canonical
Lagrangian of the SM fields and interactions. The term $h^2 R/12$ which enters
(\ref{eq:lag-eff}) with negative sign, and which was required to ensure the conformal
invariance of the original action (\ref{eq:lag-or}), will produce a small negative
contribution to the gravitational action, which will result in the eventual gravitational
coupling equal to $m_{p}^2 - v^2 / 6 \approx m_{p}^2$.

In the above model, the scalar field $\varphi$ interacts with the rest of
matter only via gravity.  Thus, the dark matter described by such a scalar would either have coexisted with the inflaton, or it would have  to
be produced, together with its perturbations, by a quantum mechanism
towards the end of the inflationary
epoch.\footnote{If we were to construct models of inflation, as done in \cite{linde1,
linde2}, we would have to provide for interaction of the inflaton field $\varphi$ with
matter fields, in order that the universe could be preheated.  This could be done,
for instance,
by replacing the constant $\lambda_h$ in (\ref{eq:lag-or}) by another conformally
invariant function $G \left( \phi / \chi \right)$ such that, at the minimum of the
potential $V (\varphi)$, the function $W (\varphi) \equiv G \left(
\tanh{\frac{\vphi}{\sqrt{6} m_p}} \right)$ acquires the value $\lambda_h$.  For the present
case, where the scalar field $\varphi$ represents dark matter, such a coupling seems to
be unnecessary, and perhaps not even desirable.}

Our primary focus in this paper will be on the following potentials belonging to the
$\alpha$-attractor family, following the prescription $V(\vphi) \to  V \left( \varphi / \sqrt{\alpha} \right)$ in (\ref{eq:pot}):
\begin{enumerate}

\item The asymmetric {\em E-Model\/} \cite{linde2}
\ber
V(\vphi) = V_0 \left ( 1 - e^{-\sqrt{\frac{2}{3 \alpha}}\frac{\vphi}{m_p}}\right )^{2n} \equiv 2V_0
\left\lbrack
\frac{\tanh{\frac{\vphi}{\sqrt{6 \alpha} m_p}}}{{1+\tanh{\frac{\vphi}{\sqrt{6 \alpha} m_p}}}} \right
\rbrack^{2n} \, . \label{eq:star}
\eer
which reduces to the potential associated with Starobinsky inflation \cite{star} when $\alpha=1$, $n=1$.

\item The tracker-potential\footnote{Note that the tracker potential
    described by (\ref{eq:sw}) is not a member of the inflationary $\alpha$-attractor
    family, which were defined to have  a  restricted dependence on $\tanh{\left(
    \frac{\vphi}{\sqrt{6\alpha}m_p}\right)}$ in order to possess a plateau. However,
    (\ref{eq:sw}) turns out to be very useful as a model of DM (and DE).}
    \cite{sahni_wang}
\ber
V(\vphi) = V_0 \sinh^{2n}\sqrt{\frac{2}{3 \alpha}}\frac{\vphi}{m_p} \equiv V_0\left\lbrack
\frac{\tanh^2\sqrt{\frac{2}{3 \alpha}}\frac{\vphi}{m_p}}
{{1-\tanh^2\sqrt{\frac{2}{3 \alpha}}\frac{\vphi}{m_p}}}
\right \rbrack^{n} \, . \label{eq:sw}
\eer

\item The {\em T-Model\/} potential \cite{linde1}
\ber
V(\vphi) = V_0 \tanh^{2n}{\frac{\vphi}{\sqrt{6\alpha}m_p}} \, . \label{eq:tanh}
\eer

\end{enumerate}

%The parameter $\lambda$ in these models plays the same role as
%$1/\sqrt{\alpha}$ in the $\alpha$-attractors of \cite{linde2}.

  An important feature of all of the above potentials is that
%for small values of $\vphi$ , all of the above potentials
they have the same asymptotic form $V(\vphi) \sim V_0 \left( \frac{\vphi}{\sqrt{6\alpha}m_p}
\right)^{2n}$ for $\frac{\vphi}{\sqrt{6\alpha}} \ll m_p$.
%$V(\vphi) \simeq \frac{1}{2}m^2\vphi^{2n}$.
It is well known that a
 scalar field oscillating about the minimum of such a potential will have
the time-averaged equation of state (EOS) \cite{turner83}
\ber
\langle w\rangle = \left\langle \frac{p}{\rho} \right\rangle = \frac{n-1}{n+1}~.
\label{eq:EOS}
\eer
Consequently, for $n=1$, the scalar field will be pressureless, just like dark matter.
For $n<1/2$, on the other hand, the EOS $\langle w\rangle <-1/3$
 violates the strong energy
condition. Such a field could therefore play the role of dark energy by
causing the universe to accelerate \cite{sahni_wang}.

\section{Dark Matter}
\label{sec:dm}

The observation that a coherently oscillating scalar field could play the role of dark
matter is not new. One of the earliest DM candidates that drew on this possibility was
the axion \cite{PQ}.  Indeed, after the global $U(1)$ Peccei--Quinn symmetry is
spontaneously broken, the pseudo-Goldstone boson of the broken symmetry, the axion,
begins to oscillate under the influence of the potential
\ber
V(\vphi) = V_0\left\lbrack 1 - \cos{\left (\frac{\vphi}{f}\right )}\right\rbrack \, .
\label{eq:axion}
\eer
In the neighbourhood of the minimum of this potential, we have $V(\vphi) \simeq
\frac{1}{2}m_a^2\vphi^2$ where $m_a^2 = V_0/f^2$ is the axion mass and $f$ is the
symmetry breaking scale. For the QCD axion, the symmetry breaking scale is experimentally
bounded by $f \geq 10^9$~GeV\@. Other phenomenological models of scalar-field dark matter
have been explored in
\cite{zeldovich,DM_early,peebles99,stein99,sahni_wang,hu00,goodman,matos00}.

An important aspect of scalar-field dark matter (SFDM) is that if the scalar field mass
is small, then the Jeans length $\lambda_J$ associated with gravitational clustering can
be very large.
%This follows from the relation \cite{zeldovich,hu00}
For the canonical massive scalar field potential
\ber
V(\vphi) = \frac{1}{2}m^2\vphi^2 \, ,
\eer
one finds \cite{zeldovich,hu00}
\ber
\lambda_J = \pi^{3/4} (G\rho)^{-1/4} m^{-1/2}~.
\label{eq:jeans}
\eer
An oscillating scalar field with a mass of $10^{-22}$~eV would therefore have a Jeans
length of a few kiloparsec. Such a large Jeans length would inhibit gravitational
clustering on small scales thereby helping to resolve the cusp--core dilemma faced by
standard cold dark matter (CDM) in the context of dwarf spheroidal galaxies
\cite{sahni_wang,hu00,Witten}. A macroscopically large Jeans length might also ameliorate
the substructure problem in CDM\@. (Warm dark matter \cite{warm} and non-canonical scalar
fields \cite{sahni_sen16} provide another means of resolving these issues.) It is of
interest to note that oscillations in the gravitational potential associated with SFDM
can induce oscillations in the photon arrival time from millisecond pulsars. This effect
may be detectable by future experiments such as the Square Kilometer Array (SKA) pulsar
timing array \cite{pulsar} and laser interferometric gravitational wave detectors
\cite{GW}. Finally, it is interesting to note that ultra-light scalar fields (pseudo
Nambu--Goldstone bosons) arise naturally in string theory via the breaking of exact shift
symmetry \cite{Witten}.

Several of  these issues have been addressed in considerable detail in \cite{Witten,fuzzy,marsh}, therefore
we do
not delve any deeper into them in this paper.
%This paper does not delve any deeper into the multifarous
% aspects of gravitational clustering
%in the scalar field scenario. This subject has been investigated in considerable
%depth in \cite{fuzzy} and the reader is referred to these papers for more detail.
Instead, our focus will be on the class of initial conditions which can give rise to
successful models of dark matter in the context of the $\alpha$-attractors. We shall
commence our study with the canonical scalar-field potential $V = \frac{1}{2} m^2\vphi^2$
and then move on to discuss the $\alpha$-attractor family of potentials
 (\ref{eq:star}), (\ref{eq:sw}) \&
(\ref{eq:tanh}).

\subsection{Dark Matter from the potential $\frac{1}{2} m^2\vphi^2$}
\label{sec:canoDM}

The canonical potential for a scalar field describing dark matter is given by
\cite{zeldovich,hu00}
\ber
V = \frac{1}{2} m^2\vphi^2~.
\label{eq:chaotic}
\eer
For small values, $\vphi \ll m_p$, the scalar field oscillates about the minimum of
its potential at $\vphi = 0$.
As discussed earlier, the averaged equation of state during oscillations is
$\langle w_\vphi\rangle \simeq 0$ which allows the scalar field to behave like dark matter.

\begin{figure}[H]
\centering
\includegraphics[width=0.7\textwidth]{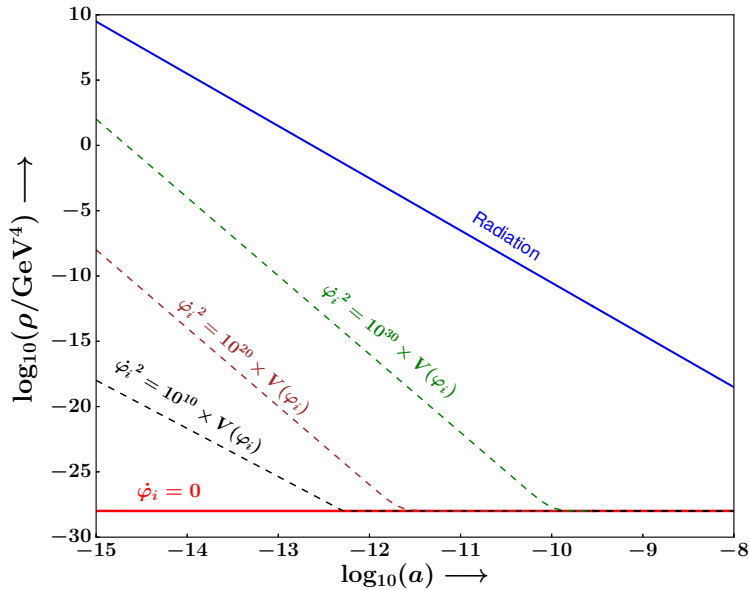}
\caption{This figure demonstrates that, for $V(\vphi)=\frac{1}{2}m^2\vphi^2$, the
scalar field velocity $\dot{\vphi}$ rapidly decays in a radiation-dominated
universe. Starting at $a=10^{-15}$ and fixing the initial field value to $\vphi_i\simeq
0.06~m_p$, we show that initially large kinetic terms rapidly get diluted, becoming
irrelevant within a few $N=\log_{10} a$ values. The black, brown and green dashed curves
show the scalar-field energy density for fixed $\varphi_i$, and (lower to higher) initial
kinetic energy values. One notes that all three curves converge towards the solid red
line corresponding to $\dot{\vphi_i}=0$ by $a\sim 10^{-10}$.    Consequently, the scalar
field starts behaving like cosmological constant by $z \sim 10^{10}$. (For clarity of
presentation, we do not show the baryon and dark energy density.) }
\label{fig:DMcan_velocity}
\end{figure}

In order to gain a deeper appreciation of this dark-matter model, we must first turn to
the radiative epoch, prior to the period when oscillations in $\vphi$
commence.\footnote{In particle-physics models of dark matter, dark matter appeared on
the cosmological scene soon after post-inflationary preheating, together
with radiation %, baryons
and other particles constituting the present universe. (Some of the particles present
after reheating might have decayed away leaving behind only a secondary relic.) In the
field-theoretic models of dark matter, such as the one under consideration in this paper,
it is conceivable that the dark-matter field could have coexisted with the Inflaton as a
spectator field, in which case it would generate a spectrum of isocurvature perturbations
\cite{liddle00}.} It is believed that the density of radiation, soon after preheating,
greatly exceeded the density of all other forms of matter in the universe. Consequently,
the expansion history at this stage can be described by the equation
\ber
H^2 = \frac{8\pi G}{3}\Bigl( \rho_r + \rho_\vphi + \rho_b + \cdots \Bigr) \simeq
\frac{8\pi G}{3}\rho_r \, ,
\eer
where
\ber
\rho_\vphi = \frac{1}{2}{\dot\vphi}^2 + V(\vphi), \qquad p_\vphi =
\frac{1}{2}{\dot\vphi}^2 - V(\vphi)~,
\eer
and the scalar field equation of motion is
\ber
{\ddot \vphi} + 3H{\dot\vphi} + V'(\vphi) = 0~.
\label{eq:eom}
\eer
Since Hubble expansion is dominated by the radiation density, the scalar field
experiences enormous damping as it attempts to roll down its potential
(\ref{eq:chaotic}). In the kinetic-dominated regime $\dot \varphi^2 \gg V
(\varphi)$, the irrelevance of $V'$ relative to the first two terms on the LHS of
(\ref{eq:eom}) ensures that the kinetic energy decreases rapidly, as ${\dot\vphi}^2
\propto a^{-6}$. This regime is then followed by slow-roll, during which the
first term in (\ref{eq:eom}) can be neglected. Consequently, in a very short span of
time, the scalar field density comes to be dominated by the potential term leading to
$\rho_\vphi \simeq V(\vphi)$. Clearly, since $\vphi$ soon virtually stops
evolving, $V(\vphi)$ begins to play the role of a cosmological constant deep in the
radiative stage. Figure~\ref{fig:DMcan_velocity} illustrates this fact.

Figure \ref{fig:DMcan_velocity} illustrates that starting from a redshift $z=10^{15}$, kinetic energy terms corresponding to  different initial velocities
$\dot{\vphi_i}$ rapidly get damped, with the result that the scalar field begins to
behave like cosmological constant by $z\simeq 10^{10}$. This is the redshift at which we
set down initial conditions in our subsequent analysis.

\begin{figure}[H]
\centering
\includegraphics[width=0.838\textwidth]{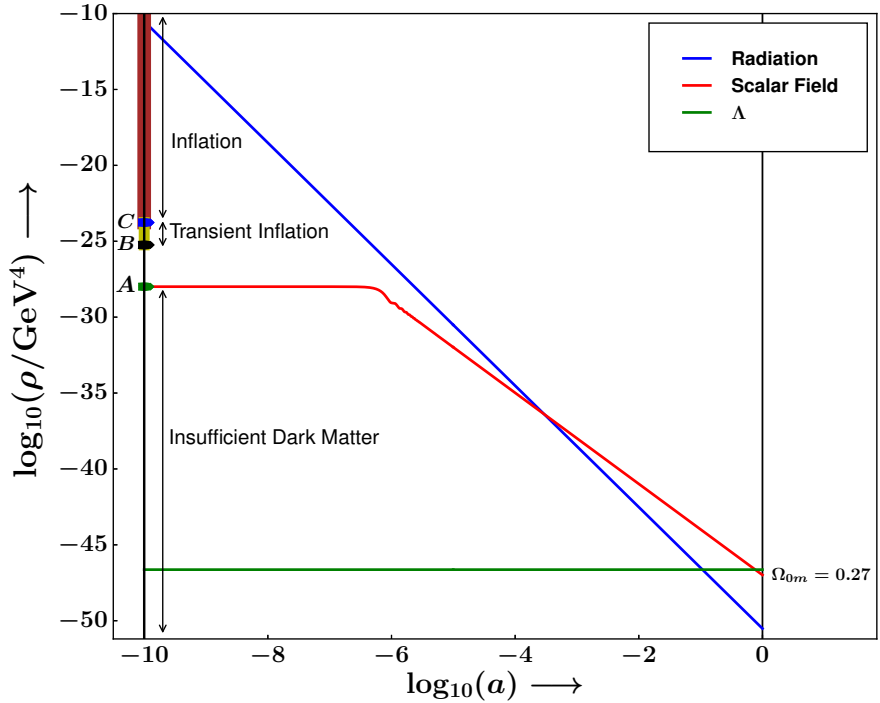}
\caption{This figure describes the fine tuning associated with the initial scalar field
density. Commencing our integration at $a=10^{-10}$ ($z \simeq 10^{10}$), we find that
the scalar-field energy density remains frozen to its initial value all the way until $z
\sim 10^6$. The rapid decline in the radiation density from $z = 10^{10}$ to $z \sim
10^6$ reduces the damping on $\vphi$ and releases the scalar field from its frozen value.
Thereafter the scalar field begins to oscillate and behave like dark matter with
$\rho_\vphi \propto a^{-3}$. The initial scalar-field value which results in $\Omega_{0m}
\simeq 0.27$ is shown by the green point $A$.  The brown color band (commencing upwards
from $C$) indicates the range of initial energy-density values which drive the universe
into an inflating (accelerating) phase that lasts until the present epoch. The narrow
green band with $\rho_i\in \big(\rho_{B},\rho_{C}\big)$ corresponds to a universe which
experiences transient acceleration. Initial values of $\rho_i < \rho_A$ result in an
insufficient amount of dark matter at the present epoch ($\Omega_{0m} < 0.27$) whereas
$\rho_A < \rho_i < \rho_{B}$ lead to too much dark matter. One, therefore, finds that,
for a given value of $m$ (in this case, $10^{-22}$\,eV), only a very narrow range
of initial values of $\vphi$ near point $A$ can lead to a current value of $\Omega_{0m}$
satisfying observational constraints. Namely, $0.057~m_p\leq \vphi_i \leq 0.062~m_p$ results in
$\Omega_{0m} = 0.27 \pm 0.03$. (For clarity of presentation, we do not show the
baryon density.)
} \label{fig:can_finetune}
\end{figure}

The tendency of $\rho_\vphi$ to behave like a cosmological constant deep within the
radiative regime enormously influences the kind of initial conditions that need to be
imposed on the scalar field in order that the model universe resemble ours (\ie,
with $\Omega_{0,\rm DE} \simeq \frac{2}{3}$, $\Omega_{0m} \simeq \frac{1}{3}$, $\Omega_{0
b} \simeq 0.04$, and $\Omega_{0r} \simeq 10^{-4}$). Figure~\ref{fig:can_finetune} illustrates
the enormous degree of fine-tuning associated with the initial density of the scalar
field.

In constructing figure \ref{fig:can_finetune}, we chose $m=10^{-22}\,{\rm eV}$. This  ensures that the
scalar field starts oscillating at $z\sim 2.8\times 10^6$, after which it begins to
behave like dark matter. For $z > 10^6$, the scalar field behaves like a cosmological
constant. As remarked earlier, scalar-field dark matter with $m \sim 10^{-22}\,{\rm eV}$
would cluster on scales greater than  a kiloparsec \cite{hu00}. Three values of the
initial scalar-field energy density, $\rho_i$, are of relevance for the present
discussion.

\noindent (i) $\rho_i = \rho_A$. This value of the initial energy density leads to a
universe just like ours; in other words, if $\rho_i = \rho_A$ then $\Omega_{0m} \simeq
0.27$. For $m = 10^{-22}\,{\rm eV}$, one finds $\rho_{A}=1.02\times 10^{-28}\,{\rm
GeV}^4$. Smaller initial values result in an insufficient amount of dark matter at the
present epoch, \ie \, $\rho_i < \rho_A \Rightarrow \Omega_{0m} < 0.27$.

\noindent (ii) $\rho_i\in \big(\rho_{B},\rho_{C}\big)$. Initial values $\rho_i \geq
\rho_B$ lead to an accelerating (inflationary) phase (sourced by the scalar field) during
the radiative epoch. For $\rho_B \leq \rho_i \leq \rho_C$, inflation is a transient, and
the universe reverts to being radiation-dominated after the scalar field has fallen to
smaller values. Larger initial values, $\rho_i > \rho_C$, result in a universe which
inflates all the way until the present. For $m = 10^{-22}\,{\rm eV}$, one finds
$\rho_{B}=5.71\times 10^{-26}\,{\rm GeV}^4$, $\rho_{C}=1.71\times 10^{-24}\,{\rm GeV}^4$.

We therefore find that the  potential (\ref{eq:chaotic}) suffers from a severe fine-tuning problem
since, for any given value of $m$, there is only a \underline{very narrow range}
of $\vphi_{i}$ which will lead to  currently admissible values of the matter
density $\Omega_{0m}$.
%since only a specific value of  initial density of scalar field  leads to a universe resembling ours.
These results support the earlier findings of \cite{stein99}.\footnote{It is interesting
to note that the fine tuning which we observe is insensitive to the initial value of
${\dot\vphi}$, as demonstrated in Figure~\ref{fig:DMcan_velocity}.}

\begin{figure}[H]
\centering
\includegraphics[width=0.72\textwidth]{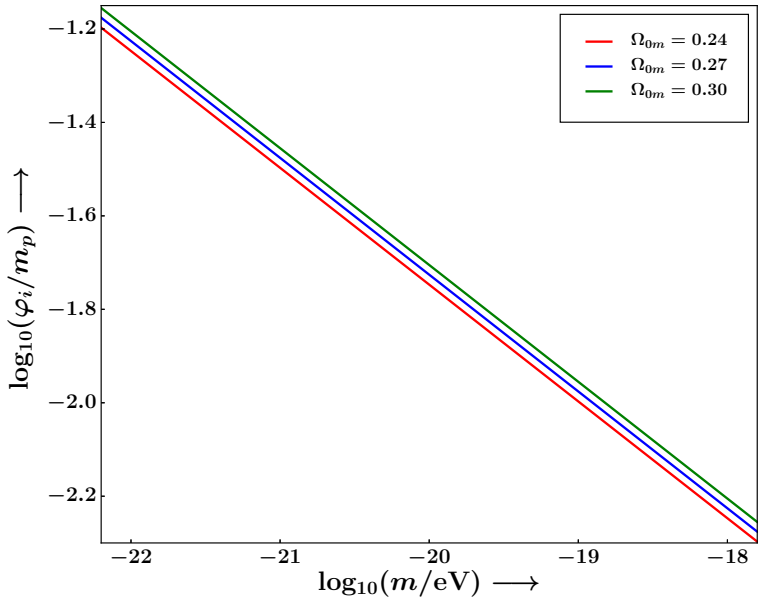}
\caption{Dependence of the initial value $\vphi_i$ of the scalar field on the mass $m$ is
shown for the scalar-field potential $V(\vphi)=\frac{1}{2}m^2\vphi^2$. Three different
values of the present matter density parameter $\Omega_{0m}$ are assumed. One sees that, for a
given value of $\Omega_{0m}$, the value of $\vphi_i$ decreases with increasing $m$
according to the relation $\vphi_i \propto m^{-{1}/{4}}$. Consequently, for a fixed value
of the mass $m$, larger values of $\Omega_{0m}$ are associated with larger initial values
of $\vphi_{i}$. } \label{fig:DMquadPM}
\end{figure}

The foregoing analysis  focused on a scalar field of mass $m=10^{-22}\,{\rm eV}$,
which presents a lower bound on the mass of an oscillating scalar field purporting to
play the role of dark matter \cite{Witten}. However, it is important to study the effect
of varying $m$ on the initial field value $\vphi_i$ and hence on the initial energy
density $\rho_i \simeq \frac{1}{2}m^2\vphi_{i}^2$. It is straightforward to show that, if
$\vphi$ plays the role of dark matter, then $\vphi_i$ scales with the mass as $\vphi_i
\propto m^{-{1}/{4}}$ (see Appendix~\ref{app:mass}). In effect,
\ber
\vphi_i=0.06\times \left(\frac{m}{10^{-22}\,{\rm eV}}\right)^{-1/4}~m_p~.
\label{eq:DMquadPM0}
\eer
Figure~\ref{fig:DMquadPM} illustrates this relationship for three different values of the
current dark-matter density parameter $\Omega_{0m}$. Note that a larger value of
$\Omega_{0m}$ requires a larger initial value $\vphi_i$ for a given scalar field mass $m$. The above
result can easily be translated into a relationship between $\rho_i$ and $m$ using
$\rho_i \simeq \frac{1}{2}m^2\vphi_{i}^2$.

\subsection{Dark Matter from the {\em E-Model\/}}
\label{sec:staroDM}

It is interesting that the extreme fine-tuning problem  faced by  the
$\frac{1}{2}m^2\vphi^2$ potential is easily alleviated if dark matter is based on the
{\em E-model}. In this case, one can write down the potential (\ref{eq:star}) with $n=1$, in the form
\ber
V(\vphi) = V_0 \left ( 1 - e^{-\lambda\frac{\vphi}{m_p}}\right )^{2}~,
\label{eq:star1}
\eer
which closely resembles the Starobinsky model  \cite{star}.

\begin{figure}[ht]
\centering
\includegraphics[width=0.69\textwidth]{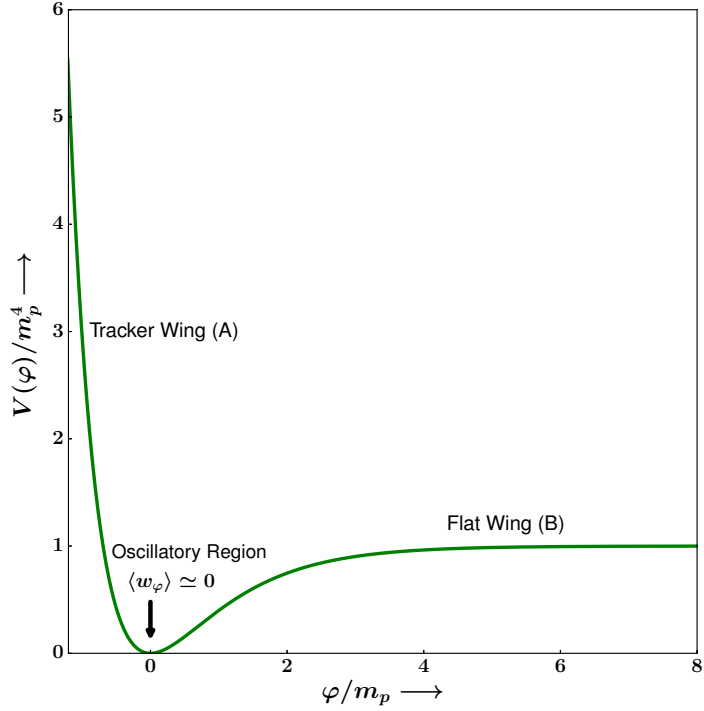}
\caption{This figure schematically illustrates the {\em E-model\/} potential (\ref{eq:star1})
 with $\lambda=1$.
The main features of this potential are: the exponential tracker wing for $\lambda|\vphi| \gg m_p$
($\vphi<0$),
the flat wing for $\lambda\vphi \gg m_p$, and the oscillatory region for which $\lambda|\vphi| \ll m_p$,
 so that
$V \simeq \frac{1}{2} m^2\vphi^2$. }
\label{fig:staroughpot}
\end{figure}

The potential (\ref{eq:star1}) exhibits three asymptotic branches (see
figure~\ref{fig:staroughpot}):
\ber
\mbox{Tracker wing:} \quad   V(\vphi) &\simeq& V_0\, e^{2\lambda|\vphi|/m_p}~, \quad
\vphi < 0\, , \quad \lambda|\vphi| \gg m_p\, ,
\label{eq:starpot1}\\
\mbox{Flat wing:} \quad V(\vphi) &\simeq& V_0\, , \quad \lambda\vphi \gg m_p \, ,
\label{eq:starpot2}\\
\mbox{Oscillatory region:} \quad V(\vphi) &\simeq& \frac{1}{2}m^2\vphi^2\, , \quad
\lambda|\vphi| \ll m_p \, , \label{eq:starpot3}
\eer
where
\ber
m^2 = \frac{2V_0\lambda^2}{m_{p}^2}.
\eer
\label{eq:starmass}
We note that the tracker parameter $\lambda$ in the potential (\ref{eq:star1}) is related to the geometric parameter $\alpha$ in (\ref{eq:star}) by 
\ber
\lambda=\sqrt{\frac{2}{3\alpha}}~.
%\frac{2}{\sqrt{6\alpha}}
\label{eq:lambda-alpha}
\eer
We now proceed to study the motion of the scalar field along the different branches (wings) of the
{\em E-model\/} potential.
\subsubsection{Evolution along the tracker wing}
\label{sec:steep}

\begin{figure}[H]
\centering
\includegraphics[width=0.85\textwidth]{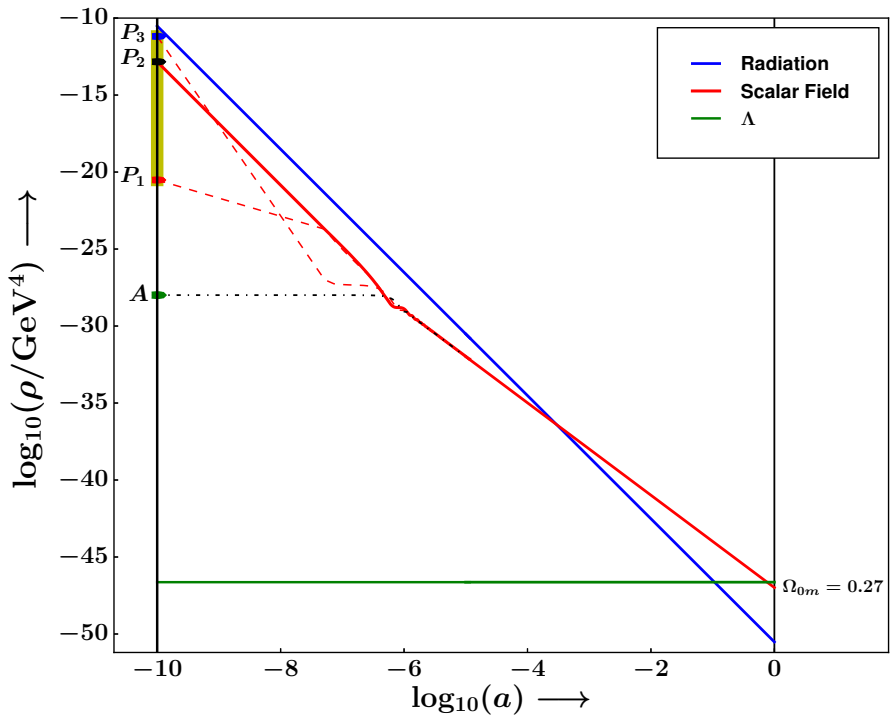}
\caption{This figure describes the evolution of the scalar-field energy density from $z
\simeq 10^{10}$ until $z = 0$. The scalar field commences its descent from the steep left
wing (`$A$' in figure~\ref{fig:staroughpot}) of the potential $V(\vphi) = V_0 \left ( 1 -
e^{-\lambda\vphi/m_p}\right )^{2} $. We have chosen $\lambda=14.5$ and $V_0=1.37\times
10^{-28}\,{\rm GeV}^4$, which correspond to $m = \frac{\sqrt{2V_0}\lambda}{m_{p}} \simeq
10^{-22}\,{\rm eV}$. The band from $P_{1}$ to $P_{3}$ represents the range in the initial
(scalar-field) energy density that leads to a reasonable value for the dark-matter
density at the present epoch, namely $\Omega_{0m} \simeq 0.27$. Point $P_{2}$
marks the initial energy density corresponding to the attractor solution (solid red
line) to which all trajectories starting in the $P_{1}$--$P_{3}$ band converge (prior to
the commencement of oscillations in $\vphi$). This behaviour is in sharp contrast to that
of dark matter sourced by the $V(\vphi)=\frac{1}{2}m^{2}\vphi^{2}$ potential, for which
only a very narrow finely tuned range of values of the initial energy density (around
point $A$) lead to $\Omega_{0m} \simeq 0.27$. (For clarity of presentation, we do not
show the baryon density.)} \label{fig:starattract}
\end{figure}

It is well known that that, in the context of Starobinsky inflation, only the flat wing
of the potential sustains inflation since the tracker wing is much too steep to cause
accelerated expansion. For dark matter, on the other hand, it is the steep wing with $V
\sim e^{2\lambda\frac{|\vphi|}{m_p}}$ that is  more useful\footnote{As mentioned earlier, the Starobinsky model of inflation corresponds to the choice $\alpha=1$ for which the left wing is not steep enough to provide tracking.}. The reason for this has to do
with the fact that a scalar field moving down a sufficiently steep potential can `track'
the cosmological background density \cite{fj97,ratra88,wang}. In the case of the
exponential potential (\ref{eq:starpot1}), if $\rho_r$ is the radiation density, then
prior to matter-radiation equality, we have
%$z \gg z_{\rm eq}$
\ber
\frac{\rho_\vphi}{\rho_{\rm total}} = \frac{1}{\lambda^2}
\label{eq:star_track}
\eer
where $\rho_{\rm total} = \rho_\vphi + \rho_r$. Thus, a field rolling down the steep wing
soon follows the common evolutionary path (\ref{eq:star_track}) from a wide range of
initial conditions.
%Thus a scalar field can start out with a large value of its energy density.
%As the universe expands, its energy remains proportional to the background
%density in the universe
This occurs so long as $\lambda|\vphi| \gg m_{p}$. The scalar field
begins to behave like dark matter, with $\langle w \rangle \simeq 0$, once
 $\vphi$
has dropped to sufficiently small values
and begins to oscillate.
The following conditions must be satisfied for this to happen \cite{sahni_wang}:
\ber
\lambda|\vphi| \ll m_p\, , \qquad m^2 \equiv V'' = \frac{2V_0\lambda^2}{m_{p}^2} =
H^2(t_*) \, .
\eer
%The behaviour described above is typical for the steep tracker wing (\ref{eq:starpot1}) and has been illustrated
This behaviour is illustrated in figure~\ref{fig:starattract} for the parameter values
$\lambda=14.5$ (corresponding to $\alpha=3.17\times 10^{-3}$) and $V_{0}=1.37\times 10^{-28}\,{\rm GeV}^{4}$. With these parameters, the
effective mass of the scalar field is $m \simeq 10^{-22}\,{\rm eV}$, and the
corresponding Jeans length is about a kiloparsec. The field begins to oscillate when $m/H
\sim 1$, which corresponds to a redshift of $z\sim 3\times 10^6$.
Figure~\ref{fig:starattract} demonstrates that the fine tuning in the initial scalar
field energy density that existed for the $m^2\vphi^2$ potential has been substantially
removed. For the sake of completeness, we give below the initial energy-density values
corresponding to:
\begin{itemize}
\item[(i)] the scaling attractor solution ($P_2$ in figure~\ref{fig:starattract}):
\ber
\rho_{P_{2}}=1.46\times 10^{-13}\, \mbox{GeV}^{4}\, ,
\eer
\item[(ii)] the maximum and minimum values of initial density that lead to
%the correct value for the current matter density
$\Omega_{0m} \simeq 0.27$:
\ber
\rho_{\rm max} (\equiv\rho_{P_3}) &=& 6.47\times 10^{-12}\,{\rm GeV}^{4}, \nonumber\\
\rho_{\rm min} (\equiv\rho_{P_1}) &=& 3.05\times 10^{-21}\,{\rm GeV}^{4}~,
\eer
which are related to the points $P_1$ and $P_3$ in figure~\ref{fig:starattract}.
\end{itemize}

Our results, summarized in figure~\ref{fig:starattract}, demonstrate that initial
energy-density values covering a range of more than 9 orders of magnitude at $z =
10^{10}$ converge onto the attractor scaling solution which gives rise to $\Omega_{0m}
\simeq 0.27$ at present. This range substantially increases if we place our initial
conditions at earlier times. For instance, if one sets $\{\vphi,\dot{\vphi}\}$ at the GUT
scale of $10^{14}\,{\rm GeV}$ ($z\sim 10^{26}$), then the range of initial density values
that converge to $\Omega_{0m}\simeq 0.27$ is an astonishing $82$ orders of magnitude\,!
The tracker branch of the {\em E-model\/} potential, therefore, allows a much greater freedom
in the choice of initial conditions than the $m^2\vphi^2$ potential. In particular it
permits the possibility of equipartition, according to which the density in dark matter
and radiation may have been comparable at very early times.

\begin{figure}[H]
\centering
\includegraphics[width=0.78\textwidth]{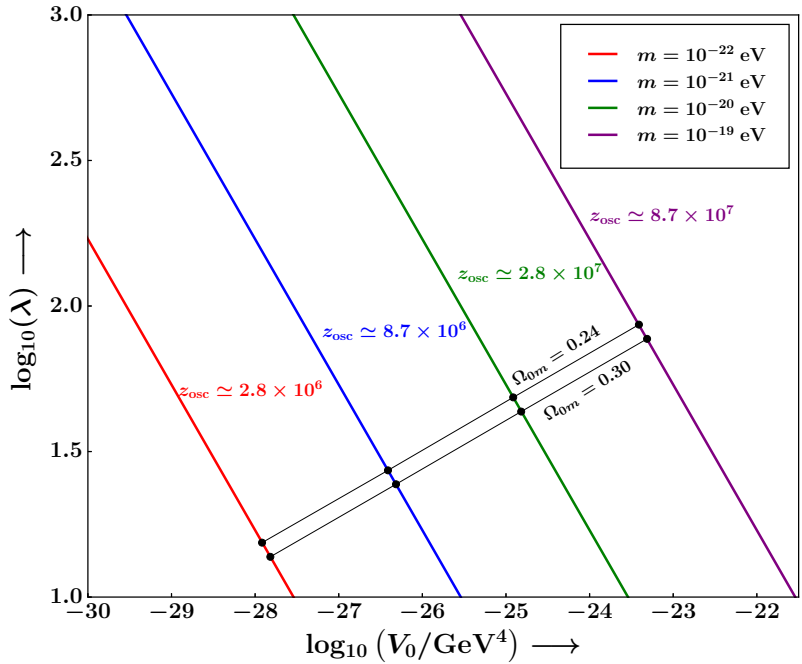}
\caption{The tracker parameter  $\lambda=\frac{2}{\sqrt{6\alpha}}$ is plotted against $V_0$ for the steep (tracker) wing of
the {\em E-model\/} potential (\ref{eq:star1}). The field begins to oscillate when $m \sim H$
where $m = \frac{\sqrt{2V_0}\lambda}{m_{p}}$. An increase in mass therefore implies an
increase in the redshift $z_{\rm osc}$ at which the scalar field begins to oscillate.
Coloured parallel lines correspond to different values of $m$ and $z_{\rm osc}$. Values
of $\lambda$ and $V_0$ which result in $\Omega_{0m}\simeq 0.24$ and $\Omega_{0m}\simeq
0.30$ are shown as parallel black lines. Their intersection with the coloured lines
describing constant values of $m$ and $z_{\rm osc}$ is marked by filled circles.
%connected by black  solid lines, on the curves describing constant values of $m$ and $z_{\rm osc}$}.
%For $\Omega_{0m}\simeq 0.3$ a relationship between $\lambda$ and $m$ accurate to within $4\%$
%within the mass range $10^{-22}\, {\rm eV} \leq m \leq 10^{-19}\, {\rm eV}$ is
%$\lambda = \left\lbrack \left (\frac{14.5^4 m}{10^{-22}\, {\em eV}}\right )^{1/2} + 1\right\rbrack^{1/2}$.
}
\label{fig:Vmeff}
\end{figure}

Our previous analysis focussed on a scalar field having an effective mass $m =
10^{-22}\,{\rm eV}$. We chose this particular value of $m$ since it could help resolve
the cusp--core and substructure problems faced by standard cold dark matter (SCDM)
\cite{sahni04,Witten}. Although masses smaller than $10^{-22}\,{\rm eV}$ could be
problematic for structure formation, larger values of $m$ are by no means ruled out.
Indeed, for $m \gg 10^{-15}\,{\rm eV}$, our model becomes indistinguishable from SCDM.
Since, in the {\em E-model\/}, $m$ is a composite quantity, being related to $\lambda$
and $V_0$ through (\ref{eq:starmass}), it is of interest to determine a general
relationship between $\lambda=\frac{2}{\sqrt{6\alpha}}$ and $V_0$ which would lead to the current value of the
matter density. Such a relationship has been plotted in figure~\ref{fig:Vmeff}, in which
values of $V_0$ and $\lambda$ giving rise to $\Omega_{0m} = 0.24$ and $0.30$ are shown as
black lines. One should note that each point on either of these two lines refers to a
family of initial conditions $\{\vphi_i,\dot{\vphi_i}\}$ which get funneled onto a given
$\Omega_{0m}$ by means of the attractor mechanism described in
figure~\ref{fig:starattract}.

\subsubsection{Evolution along the flat wing}
\label{sec:flat}

\begin{figure}[H]
\centering
\includegraphics[width=0.735\textwidth]{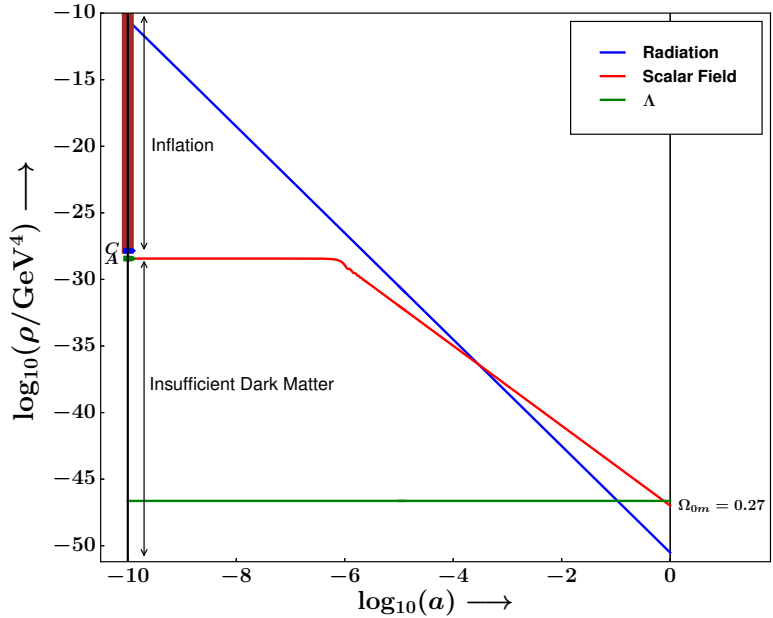}
\caption{This figure describes the possibility that the scalar field associated with dark
matter could have commenced rolling from the flat right wing (see
figure~\ref{fig:staroughpot}) of the  potential $V(\vphi) = V_0 \left ( 1 -
e^{-\lambda\frac{\vphi}{m_p}}\right )^{2}$. We assume the same units as in
figure~\ref{fig:starattract}, namely  $\lambda=14.5$ and $V_0=1.37\times 10^{-28}\,{\rm
GeV}^4$ which correspond to $m \simeq 10^{-22}\,{\rm eV}$. The energy densities
(in units of ${\rm GeV}^{4}$) of radiation and scalar field are plotted against the scale
factor $a$. We notice that the scalar-field energy density remains pegged to its initial
value when the radiation energy density is large. Subsequently, the radiation energy
density drops, which decreases the damping in the scalar-field equation of motion
(\ref{eq:eom}). From this point onwards ($z \sim 10^6$), the scalar field is free to move
and to oscillate. Consequently, from here on it behaves like dark matter. Point $A$
describes the (fine-tuned) initial value of the scalar-field energy density which leads
to $\Omega_{0m}\simeq 0.27$. The brown band pointing upwards from point $C$ shows the
range of initial values which drive the universe into an inflationary accelerating phase
which lasts until the present epoch. Small initial values $\rho_i \ll \rho_A$ lead to an
insufficient amount of dark matter at present. (For clarity of presentation, we do not
show the baryon density.)} \label{fig:starRfinetune}
\end{figure}

Next, we explore  the possibility of obtaining dark matter from the flat right wing of
the {\em E-model\/} potential (\ref{eq:starpot2}). (It is this wing which is responsible for
generating inflation in the Einstein frame of the Starobinsky model.) Note that the value
of $V_0$ sets the height of the flat wing of the potential in
figure~\ref{fig:staroughpot}. The fact that the effective mass of the scalar field
depends upon $V_0$ through $m = {\sqrt{2V_0}\lambda}/{m_{p}}$ opens up several different
possibilities for initial values of the scalar field. For instance, if $m$ is held fixed
at $10^{-22}\,{\rm eV}$ and the values of $\lambda$ and $V_0$ are chosen in conformity
with our previous analysis, namely $\lambda=14.5$ and $V_0=1.37\times 10^{-28}\,{\rm
GeV}^4$, then
%We first describe the situation using the same parameters with which we explored the steep right wing.
our results, shown in figure~\ref{fig:starRfinetune}, demonstrate that, as in the case of
the $m^2\vphi^2$ potential, one encounters here a considerable fine tuning of initial
conditions.
For instance, the fine-tuned initial value of the scalar field density which ensures
$\Omega_{0m}=0.27$, is given by $\rho_A=3.68\times 10^{-29}\,{\rm GeV}^{4}$. If the
initial energy density of the  scalar field is larger than $\rho_B =1.32\times
10^{-28}\,{\rm GeV}^{4}$, then the universe enters into an accelerating (inflationary)
phase. For values of the initial energy density between $\rho_B$ and $\rho_C=1.37\times
10^{-28}\,{\rm GeV}^{4}$, the accelerating phase is a transient, whereas for initial
energy-density values larger than $\rho_C=1.37\times 10^{-28}\,{\rm ~GeV}^{4}$ (brown
band in figure~\ref{fig:starRfinetune}) the inflationary phase lasts until the present
epoch. As in the case of the $m^2\vphi^2$ potential, initial energy-density values
smaller than $\rho_A$ result in an insufficient amount of dark matter at the current
epoch. (Note that the values of $\rho_B$ and $\rho_C$ are much
 too close to be distinguished in figure~\ref{fig:starRfinetune}.)
We therefore conclude that the flat right wing of the {\em E-model\/} potential with
$\lambda=14.5$, $V_{0}=1.37\times 10^{-28}\,{\rm GeV}^{4}$ and $m=10^{-22}\,{\rm eV}$
suffers from a fine-tuning problem similar to the one which afflicts the $m^2\vphi^2$
potential.

%We also draw attention to the fact that % in the scenario described in fig. \ref{fig:starRfinetune}
%for this value of $V_0$ the value of $\vphi_{i}$
%(corresponding to $\rho_A$ in fig. \ref{fig:starRfinetune}) is not large enough to place the field onto
%the flat right region of the potential where $V' \simeq 0$.
%Instead, $\vphi$ commences its descent from a somewhat lower region in the potential
%where $V'>0$.
%One therefore encounters a situation similar to that encountered earlier
% for the $m^2\vphi^2$ potential. As in that case,
%the value of $\dot{\vphi_i}$ doesn't play a significant role in the overall scenario; see figure~\ref{fig:DMcan_velocity}.
%However this is not the case if $V_0$ is decreased, since $\vphi_i$ can now commence rolling from on top of
% and $\vphi_i$ is not too large,
%  as shown in figure \ref{fig:flat1}.
%(\ref{eq:starmass})
 %However the initial density values $\rho_A$, $\rho_B$ and  $\rho_C$ are different from the values corresponding to the quadratic potential as the two potentials have different mathematical forms.
\begin{figure}[H]
\centering
\includegraphics[width=0.717\textwidth]{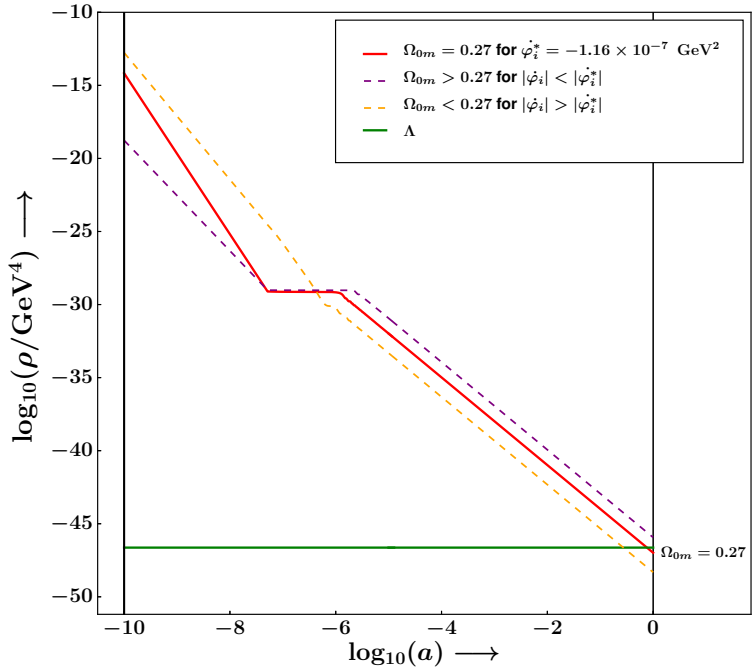}
\caption{This figure illustrates the importance of the initial velocity $\dot{\vphi_i}$
in determining the extent of dark matter at the present epoch. Again, the scalar field
rolls down the flat right wing of the {\em E-model\/} potential (\ref{eq:star1}). The height
of the potential is now somewhat lower than that in figure~\ref{fig:starRfinetune},
namely $V_0=10^{-29}\,{\rm GeV}^{4}$. Again, $m = 10^{-22}\,{\rm eV}$ with $\lambda$
determined from (\ref{eq:starmass}). The comparatively lower value of $V_0$ allows the
field to commence rolling from on top of the flat potential where $V'\simeq 0$. One finds
that, for the initial value $\vphi_i=7.35\times 10^{-2}m_p$, only a highly
fine-tuned value of $\dot{\vphi_i}=\dot{\vphi_{i}}^{*}$ results in the correct value of
the dark-matter density today, namely $\Omega_{0m}\simeq 0.27$ (red curve).
Slight deviations from this value of $\dot{\vphi_i}$ result either in an
overdensity of $\Omega_{0m}>0.27$ (purple dashed line) or in an underdensity
$\Omega_{0m}<0.27$ (orange dashed line). (For clarity of presentation, we do not show the
radiation and baryon energy density.)} \label{fig:flat1}
\end{figure}

Note that the location of the initial value $\vphi_i$ on the flat wing can be crucial in
determining the fate of the universe.
%the flat right wing where $V' \simeq 0$. In this case the value of the
%initial velocity $\dot{\vphi_i}$ becomes important when analysing the dynamics of $\vphi(t)$.
For instance, if $\vphi_i$ is large enough to place the field on top of the flat wing
where $V' \simeq 0$, then
%This is easy to see by noting that for a scalar field rolling
% along the flat wing,
the equation of motion (\ref{eq:eom}) simplifies to
${\ddot\vphi} + 3H{\dot\vphi} \simeq 0$, which has the solution ${\dot\vphi} \propto a^{-3}$.
%Thus unless ${\dot\vphi_{i}^2}$ is very large,
Thus $\vphi$ can grind to a halt en-route to the minimum of the potential.
In this case, if the initial value $\vphi_i$ is large and if ${\dot\vphi_{i}^2} \ll V_0$,
then the kinetic term will decay completely while $\vphi$ is still on the flat wing.
This can lead to eternal inflation.
%At the very least this will delay the onset of oscillations and result in
This problem can be avoided if $\vphi_i$ is not too large,
and if $({\dot\vphi^2})_i \gg V_0$.
The effect of varying the initial scalar-field velocity $\dot{\vphi_i}$ on the current
dark-matter density is shown in figure~\ref{fig:flat1}. We start with a moderate initial
value $\vphi_i=7.35\times 10^{-2}m_p$, which ensures that the field does not commence
rolling from too far along the flat wing, and thereby avoids the problem of eternal
inflation. We then vary $\dot{\vphi_i}$ for a potential with $V_0=10^{-29}\,{\rm GeV}^4$.
(The corresponding value of $\lambda$ is determined from (\ref{eq:starmass}) assuming $m=
10^{-22}\,{\rm eV}$.)  The  value of $\dot{\vphi_i}$ which gives rise to
$\Omega_{0m}=0.27$ is given by $\dot{\vphi_i}^{*}=-1.16\times 10^{-7}\,{\rm GeV}^2$ (red
line). Smaller values of $\dot{\vphi_i}$ delay the onset of oscillations in the scalar
field. This leads to a larger value of the dark-matter density at the present epoch,
$\Omega_{0m}>0.27$ (purple dashed line). A larger initial kinetic term has the opposite
effect of making the scalar field oscillate earlier, which results in a smaller value of
the dark-matter density, $\Omega_{0m}<0.27$ (orange dashed line).
\begin{figure}[H]
\centering
\includegraphics[width=0.67\textwidth]{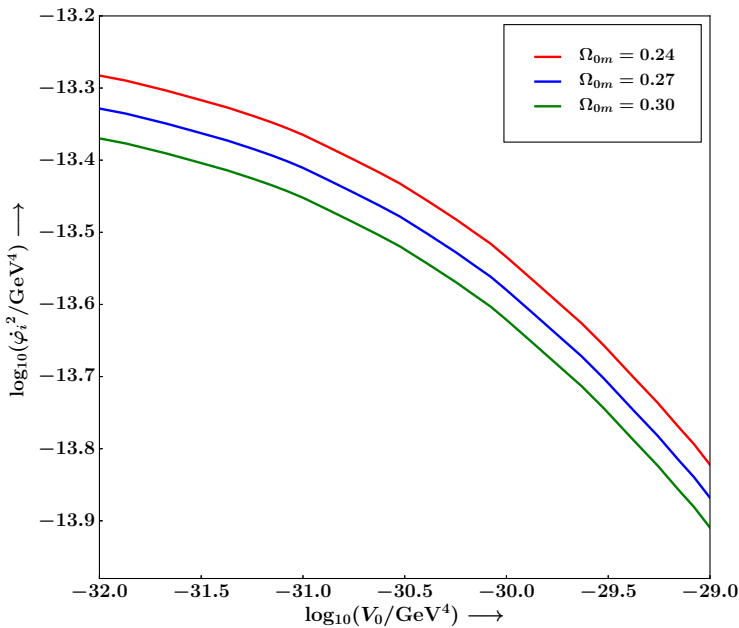}
\caption{Values of $\dot{\vphi_i}^2$ and $V_0$ resulting in $\Omega_{0m}\simeq 0.24$,
$0.27$, and $0.30$ are shown. Note that $V_0$ describes the height of the flat wing of
the potential in (\ref{fig:staroughpot}), and that larger values of $\dot{\vphi_i}^2$
correlate with smaller values of $V_0$ for a given $\Omega_{0m}$. } \label{fig:flat2}
\end{figure}

\begin{figure}[H]
\centering
\includegraphics[width=0.665\textwidth]{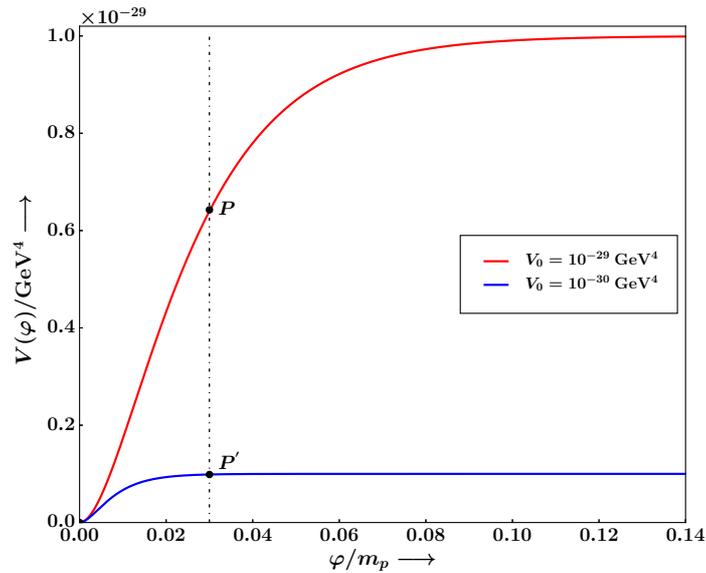}
\caption{The right wing of the {\em E-model} potential $V(\vphi)$ is shown for two values
of $V_0$. % satisfying the relation  $m = \frac{\sqrt{2V_0}\lambda}{m_{p}}=10^{-22}\,{\rm eV}$.
%Notice that for smaller values of $V_0$, the same initial value $\vphi_i=0.03m_p$
Notice that, for a lower potential (blue), the initial value $\vphi_i$ is associated with
a point $P'$ located along the {\em flat wing\/} of $V(\vphi)$. By contrast, the same
value of $\vphi_i$ is associated with the {\em steep wing\/} $P$ of the higher potential
(red).} \label{fig:flat3}
\end{figure}

The  dependence of the initial kinetic term $\dot{\vphi_i}^2$ on the height $V_0$ of the
flat wing has been numerically analyzed  with the initial field value held fixed at
$\vphi_i=7.35\times 10^{-2}m_p$. Our results are shown in figure~\ref{fig:flat2}. We find
that smaller values of $V_0$ require larger values of $\dot{\vphi_i}^2$ in order to give
the same final value of $\Omega_{0m}$. This correlation is expected since one obeys the
mass relation (\ref{eq:starmass}) while varying the height $V_0$ of the potential. For
smaller $V_0$ values, an initial $\vphi_i=7.35\times 10^{-2}m_p$ sits {\em further
along\/} on the flat wing of the potential (see figure~\ref{fig:flat3}). Hence the field
requires a larger initial velocity $\dot{\vphi_i}$ in order to start oscillating at the
correct epoch ($z_{\rm osc}\sim 10^{6}$ for $m= 10^{-22}\,{\rm eV}$) thereby ensuring
$\Omega_{0m}=0.27$.

\begin{figure}[H]
\centering
\includegraphics[width=0.66\textwidth]{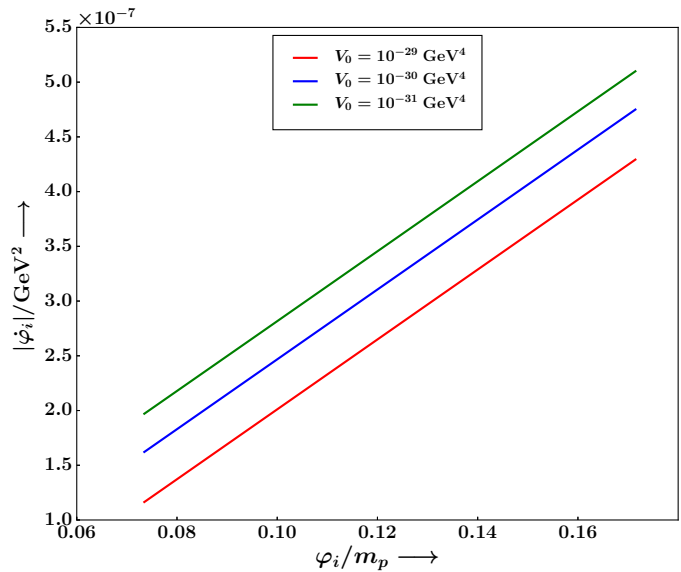}
\caption{The initial velocity $\dot{\vphi_i}$ which results in $\Omega_{0m}=0.27$ is
plotted as a function of $\vphi_i$ for three different values of the potential height
$V_0$. Notice that, for a given $V_0$, a larger value of $\vphi_i$ implies a larger
$|\dot{\vphi_i}|$. However, if $\vphi_i$ is held fixed, then $|\dot{\vphi_i}|$ decreases
with increasing $V_0$, in agreement with figure~\ref{fig:flat2}.} \label{fig:flat4}
\end{figure}

As might be expected, for a fixed value of $V_0$, the initial velocity $\dot{\vphi_i}$
required to achieve a given value of $\Omega_{0m}$ increases with an increase in the
 initial field value $\vphi_i$.\footnote{We only consider fields
rolling towards the minimum of the potential, since a field moving in the opposite direction will give rise to
eternal inflation for all values of $\dot{\vphi_i}$.} This is illustrated in
 figure~\ref{fig:flat4} which also shows
 that the dependence of  $\dot{\vphi_i}$ on $\vphi_i$ is linear.
\subsection{Dark Matter from the Tracker Model}
\label{sec:TracDM}

For $n=1$, the tracker potential (\ref{eq:sw}) can be written as
\ber
V(\vphi) = V_0 \sinh^{2}\frac{\lambda\vphi}{m_p}~,
\label{eq:sw1}
\eer
where the parameter $\lambda$ is related to the geometric parameter $\alpha$ in (\ref{eq:sw}) by
\ber
\lambda=\sqrt{\frac{2}{3\alpha}}~.
%\frac{2}{\sqrt{6\alpha}}
\label{eq:lambda-alpha1}
\eer

\begin{figure}[H]
\centering
\includegraphics[width=0.68\textwidth]{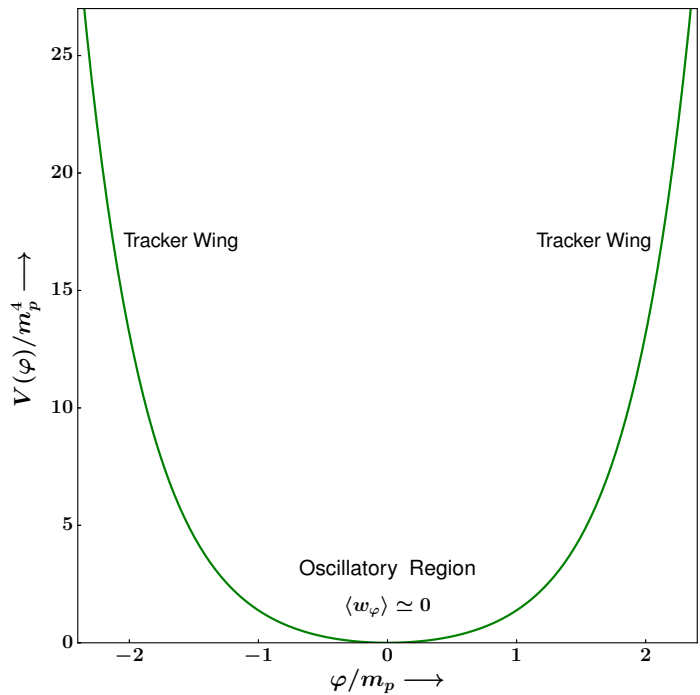}
\caption{This figure schematically illustrates the tracker potential (\ref{eq:sw1}).
The asymptotic tracker wings correspond to $\lambda|\vphi| \gg m_p$,
while, for $\lambda|\vphi| \ll m_p$, we have $V(\vphi) \simeq \frac{1}{2}m^2\vphi^2.$}
\label{fig:Trac}
\end{figure}

This symmetric potential has the following branches (see
figure~\ref{fig:Trac}):
\ber
\mbox{Two tracker wings:} \quad   V(\vphi) &\simeq& V_0\,
e^{2\frac{\lambda|\vphi|}{m_p}}~, \quad \lambda|\vphi| \gg m_p \, ,
\label{eq:sw3}\\
\mbox{Oscillatory region:} \quad V(\vphi) &\simeq& \frac{1}{2}m^2\vphi^2\, , \quad
\lambda|\vphi| \ll m_p \, , \label{eq:sw4}
\eer
where $m^2 = 2V_0\lambda^2/m_{p}^2$ and the relation between the tracker parameter $\lambda$ and  the geometric parameter $\alpha$ is given by (\ref{eq:lambda-alpha}).

\begin{figure}[H]
\centering
\includegraphics[width=0.85\textwidth]{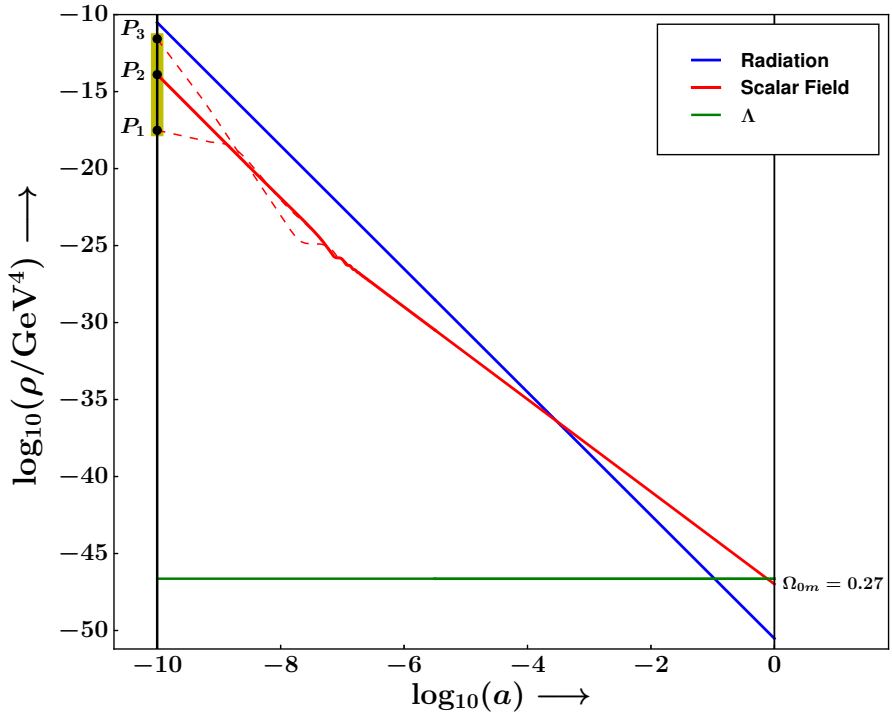}
\caption{This figure describes the evolution of the density in a scalar field which
commences its descent from one of the  steep tracker wings of $V(\vphi) = V_0
\sinh^{2}\frac{\lambda\vphi}{m_p} $ at $z \simeq 10^{10}$.
%until $z = 0$.
%The scalar field commences its descent from either of the  steep tracker wings (shown in figure~\ref{fig:Trac}) of the
%potential $V(\vphi) = V_0 \sinh^{2}\frac{\lambda\vphi}{m_p} $.
We choose $\lambda=\frac{2}{\sqrt{6\alpha}}=48.9$  and $V_0=1.2\times 10^{-25}\,{\rm GeV}^4$ so that $m =
\frac{\sqrt{2V_0}\lambda}{m_{p}} \simeq 10^{-20}\,{\rm eV}$. The band from $P_{1}$ to
$P_{3}$ represents  the range in the initial energy-density values which converge to $\Omega_{0m}
\simeq 0.27$ at present. $P_{2}$ denotes the initial density corresponding to the
attractor solution (solid red line) to which all trajectories starting on the
$P_{1}$--$P_{3}$ band converge prior to the commencement of oscillations. (For clarity of
presentation, we do not show the baryon density.)}
 \label{fig:DMtrac}
\end{figure}

The potential (\ref{eq:sw1}) converges to an  exponential for large values of
$\lambda|\vphi|$. It therefore has much in common with the left wing of the {\em E-model}
potential (\ref{eq:star}). The evolution of energy densities in radiation, a cosmological
constant and the scalar field is shown in figure~\ref{fig:DMtrac} for parameter values
$\lambda=48.93$ (corresponding to $\alpha=2.78\times10^{-4}$) and $V_0=1.20\times 10^{-25}\,{\rm GeV}^4$, which correspond to $m =
10^{-20}\,{\rm eV}$. For this value of $m$, the scalar field begins to oscillate at
$z\sim 3\times 10^7$.

The  tracker-like properties of this potential, shown in
figure~\ref{fig:DMtrac}, ensure that  initial density values covering a
range of almost 6 orders of magnitude (at $z=10^{10}$),
converge onto the attractor scaling solution which gives  $\Omega_{0m}\simeq
0.27$. This range covers more than 73 orders of magnitude if one sets  initial
conditions at the GUT scale of $10^{14}\,{\rm GeV}$ ($z\sim 10^{26}$).

The initial energy-density value corresponding to the scaling attractor solution $P_2$ in
figure~\ref{fig:DMtrac} is $\rho_{P_{2}}=1.28\times 10^{-14}\,{\rm GeV}^{4}$.
The maximum and minimum values of the initial energy density that lead to
%the correct value for the current matter density
$\Omega_{0m} \simeq 0.27$ are given by
\ber
\rho_{\rm max} (\equiv\rho_{P_3}) &=& 2.65\times 10^{-12}\, {\rm GeV}^{4}\, , \nonumber\\
\rho_{\rm min} (\equiv\rho_{P_1}) &=& 3.05\times 10^{-18}\, {\rm GeV}^{4}~.
\eer
These are related, respectively, to the points $P_1$ and $P_3$ in
figure~\ref{fig:DMtrac}.

The relationship between $\lambda$ and $V_0$ is similar to that shown in figure \ref{fig:Vmeff} for the steep wing of the {\em E-model\/}, therefore we do not show it here. Because of the presence of symmetric tracker wings, this model is perhaps the most robust
of the three $\alpha$-attractor potentials we have considered in providing a scalar-field
description of dark matter.
\subsection{Dark Matter from the T-Model}
\label{sec:TDM}

\begin{figure}[H]
\centering
\includegraphics[width=0.67\textwidth]{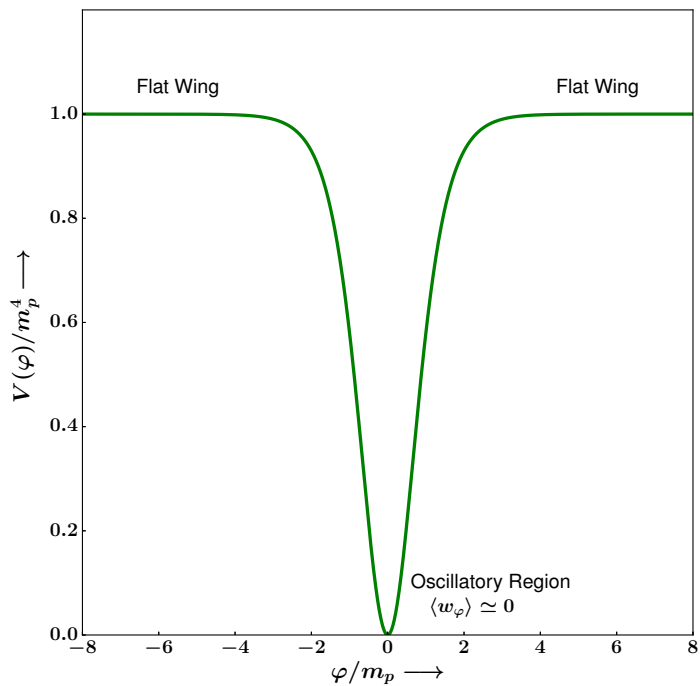}
\caption{This figure schematically shows the T-model potential (\ref{eq:tanh1}).
This potential is characterized by two asymptotically flat wings (for $\lambda_1|\vphi| \gg m_p$)
and an intermediate region in which $V \sim m^2\vphi^2$ (for $\lambda_1|\vphi| \ll m_p$).
}
\label{fig:T}
\end{figure}

The T-model potential (\ref{eq:tanh})
with $n=1$  becomes
%can provide an example of scalar field dark matter. In this case
%the potential
\ber
V(\vphi) = V_0 \tanh^{2}{\lambda_1\frac{\vphi}{m_p}}~
\label{eq:tanh1}
\eer
where the potential parameter $\lambda_1$ is related to the geometric parameter $\alpha$ by 
\ber
\lambda_1=\frac{1}{\sqrt{6\alpha}}
\label{eq:lambda-alpha2}
\eer
This potential is characterized by two asymptotically flat regions
and an intermediate region in which, for $\lambda_1|\vphi| \ll 1$, the potential behaves like
 $m^2\vphi^2$ (see figure~\ref{fig:T}). In other words,
\ber
{\rm for} ~\lambda_1|\vphi| \gg m_p \, , \quad V(\vphi) &\simeq& V_0\, , \quad \mbox{(flat
wings)} \, , \nonumber\\
%\label{eq:flat}
{\rm for} ~\lambda_1|\vphi| \ll m_p \, , \quad V(\vphi) &\simeq& \frac{1}{2}m^2\vphi^2\, ,
\label{eq:non-flat}
\eer
where $m^2 = {2V_0\lambda_{1}^{2}}/{m_{p}^2}$.

Clearly, the flat wings of this potential are very similar to the flat right wing of the
{\em E-model\/} potential (\ref{eq:star1}). Consequently the analysis for this model is
qualitatively similar to that done in section \ref{sec:flat}. As in that case, the
initial velocity $\dot{\vphi_i}$ plays an important role in determining the value of the
current matter density $\Omega_{0m}$. We do not repeat this analysis here, referring
instead the reader to figures \ref{fig:starRfinetune}, \ref{fig:flat1}, \ref{fig:flat2},
\ref{fig:flat3} and  \ref{fig:flat4}   for details. One might note that, because of the
absence of an exponential tracker wing in ({\ref{eq:tanh1}}), the initial conditions
which give rise to a realistic construct for dark matter in the T-model are less generic
than those discussed in the previous section \ref{sec:TracDM} for the tracker model.

\subsection{Gravitational Instability}
\label{sec:grav}

We demonstrated in the previous section that dark matter based on $\alpha$-attractors
could emerge from a much larger class of initial conditions than admitted by the
canonical model $\frac{1}{2}m^2\vphi^2$.
It is well known that a coherently oscillating scalar field experiences
gravitational instability \cite{Branden}.
In the case of the canonical model the associated Jeans wavenumber given by \cite{zeldovich,hu00}
\ber
k_J^2 = \sqrt{2\rho}\,\frac{m}{m_p}~.
%k_J^2 = 4\sqrt{\pi G\rho}\,m~.
%\lambda_J = \pi^{3/4} (G\rho)^{-1/4} m^{-1/2}~.
\label{eq:jeans0}
\eer
For small values of $m$, the large Jeans length associated with ultra-light (fuzzy) dark
matter might help resolve several shortcomings of standard CDM including the cusp--core
problem, the substructure problem, etc.\@ (see \cite{Witten} for a recent review). It
therefore becomes important to determine the Jeans length associated with the
$\alpha$-attractor family of dark-matter models (\ref{eq:star1}), (\ref{eq:sw1}) and
(\ref{eq:tanh1}). This shall be the purpose of the present section.
%(\ref{eq:jeans}).
%Since all $\alpha$-attractors contain $\frac{1}{2}m^2\vphi^2$ as a limiting case
%when $\vphi \to 0$, it might appear at first glance, that the late time
%properties of the $\alpha$-attractor family are
% identical. This, however, is not the case.
%In this section we demonstrate that the clustering properties of dark matter
%can vary significantly between the different $\alpha$-attractors
%(\ref{eq:tanh1}), (\ref{eq:star1}),
%and (\ref{eq:sw1}).

One first notes that, for small values of $\vphi$, the potentials corresponding to
 tracker dark matter (\ref{eq:sw1}), and the T-model (\ref{eq:tanh1})
acquire a form similar to that of the anharmonic oscillator
\ber
V(\vphi) = \frac{1}{2}m^2\vphi^2 + \frac{1}{4}{\lambda_0 }\vphi^4~. \label{eq:anharmonic}
\eer
%\ber
% m^2 = 2V_0\lambda^2, ~~{\lambda_0 } = -\left (\frac{2}{3}\right )^3 V_0\lambda^4
%\label{eq:tmodel2}
%\eer
%describing the T-model (\ref{eq:tanh1}).
Gravitational instability in such a potential %the anharmonic potential (\ref{eq:anharmonic})
was analyzed in \cite{zeldovich,kamion} in the limit $\lambda_0 \vphi^2\ll m^2$.
Employing the theory of parametric resonance (see Appendix~\ref{app:instab} for details),
one obtains the following expression for the Jeans scale ($k_J = 2\pi/\lambda_J$):
\ber
k_J^2 = -\frac{9}{8}{\lambda_0 }\vphi_0^2 + \sqrt{\left (\frac{9}{8}{\lambda_0
}\vphi_0^2\right )^2 + \frac{m^4 \vphi_0^2}{m_{p}^2}}~,
\label{eq:jeans scale}
\eer
equivalently
\ber
k_J^2 = -\frac{9}{4} {\lambda_0 } \frac{\rho}{m^2} + \sqrt{\left( \frac{9}{4} {\lambda_0
} \frac{\rho}{m^2} \right)^2  + 2 \rho \frac{m^2}{m_p^2} }~,
\label{eq:jeans_anharmonic}
\eer
where one notes that $\rho = \frac{1}{2} m^2\vphi_0^2$, with $\vphi_0$ being the maximum
amplitude of an oscillation. As expected, (\ref{eq:jeans scale}) reduces to
(\ref{eq:jeans0}) for ${\lambda_0 } = 0$.

For small values, $\lambda_1|\vphi| \ll m_p$, the T-model potential (\ref{eq:tanh1}) reduces to
\ber
V(\vphi) \simeq V_0 \left[ \left( \lambda_1\frac{\vphi}{m_p}\right)^2 - \frac{2}{3} \left(
\lambda_1 \frac{\vphi}{m_p} \right)^4 \right] \, . \label{eq:tmodel_asymp}
\eer
Comparing (\ref{eq:tmodel_asymp}) with (\ref{eq:anharmonic}) one finds
\ber
 m^2 = \frac{2V_0\lambda_1^2}{m_{p}^2}\, , \qquad {\lambda_0 } = -\frac{8}{3} \frac{V_0\lambda_1^4}{m_{p}^4}~.
\label{eq:tmodel2}
\eer
We note that $\lambda_1$ is related to $\alpha$ by (\ref{eq:lambda-alpha2}).
Substituting (\ref{eq:tmodel2}) into (\ref{eq:jeans scale})
and using $\rho = \frac{1}{2} m^2\vphi^2$, gives the Jeans scale in the T-model (\ref{eq:tanh1})
\ber
k_J^2 = 3\frac{\lambda_1^2\rho}{m_{p}^2} + \left \lbrack \left
(3\frac{\lambda_1^2\rho}{m_{p}^2}\right )^2 + 2\rho \frac{m^2}{m_{p}^2}\right
\rbrack^{1/2}~.
 \label{eq:jeans_Tmodel}
\eer
Note that $\lambda_1$ is not a free parameter since it enters
into the definition of $m^2$ in (\ref{eq:tmodel2}).

%It is interesting to note that the expression for $\lambda_J$ in (\ref{eq:jeans1})
%is very similar to that in the axion model (\ref{eq:axion}). This can easily be
%seen by expanding the axion potential in (\ref{eq:axion}) in terms of $\vphi$
%as follows .....
For small values of $\vphi$ the axion potential (\ref{eq:axion}) also has the anharmonic
form (\ref{eq:anharmonic}) with
\ber
m^2=\frac{V_0}{f^2}\, , \qquad {\lambda_0 } = -\frac{1}{6}\frac{m^2}{f^2} \, .
\label{eq:axion_jeans}
\eer
Substituting this in (\ref{eq:jeans scale}) gives the Jeans scale in the axion model
\ber
k_J^2 = \frac{3\rho}{8f^2} + \left\lbrack \left (\frac{3\rho}{8f^2}\right )^2 +2\rho
\frac{m^2}{m_{p}^2}\right \rbrack^{1/2}~.
\label{eq:jeans_axion}
\eer
Comparing (\ref{eq:jeans_axion}) with (\ref{eq:jeans_Tmodel}), one finds a close affinity
between the axion and the T-model, with (\ref{eq:jeans_axion}) reverting to
(\ref{eq:jeans_Tmodel}) after setting $1/8f^2 = \lambda_1^2$; also see \cite{kamion}.

Next consider tracker dark matter. The potential (\ref{eq:sw1}) in the limit $\lambda\vphi \ll 1$
takes the form
\ber
V(\vphi) \simeq V_0\left\lbrack \left (\frac{\lambda\vphi}{m_{p}}\right )^2 + \frac{1}{3}\left (\frac{\lambda\vphi}{m_{p}}\right )^4\right\rbrack~.
\label{eq:sw_asymp}
\eer
Comparing with (\ref{eq:anharmonic}), one finds
\ber
m^2 = \frac{2V_0\lambda^2}{m_{p}^2}\, , \qquad {\lambda_0} =
\frac{4}{3}\frac{V_0\lambda^4}{m_{p}^4}\, , \label{eq:sw_jeans}
\eer
and substitution in (\ref{eq:jeans scale}) gives the Jeans scale in the Tracker model (\ref{eq:sw1})
\ber
k_J^2 = - \frac32 \lambda^2\rho +  \left\lbrack \left (\frac32 \lambda^2\rho\right )^2 +
\rho \frac{m^2}{m_{p}^2}\right \rbrack^{1/2}~,
\label{eq:jeans3}
\eer
where $\rho = \frac{1}{2}m^2\vphi_0^2$. Note that $\lambda$ is not a free parameter since it enters
into the definition of $m^2$ in (\ref{eq:sw_jeans}).

Finally, let us consider the small argument limit of the {\em E-model\/} potential
(\ref{eq:star1})
\ber
V(\vphi) \simeq V_0\left\lbrack \left (\frac{\lambda\vphi}{m_{p}}\right )^2 - \left (\frac{\lambda\vphi}{m_{p}}\right
)^3 + \frac{7}{12}\left (\frac{\lambda\vphi}{m_{p}}\right )^4\right\rbrack ~.
\label{eq:star_asymp}
\eer
This  potential does not subscribe to the form (\ref{eq:anharmonic}) since it contains a
cubic term in $\vphi$. As shown in Appendix~\ref{app:instab}, the Jeans scale in this
model is given by
\ber
k_J^2 = \left(\frac{5}{3}\frac{\mu^2\rho}{m^4}-\frac{9}{4}\frac{{\lambda_0
}\rho}{m^2}\right)+\left[\frac{2m^2}{m_{p}^2}\rho+\left(\frac{25}{9}\frac{\mu^4}{m^8}+\frac{81}{16}\frac{{\lambda_0
}^2}{m^4}-\frac{15}{2}\frac{{\lambda_0 }\mu^2}{m^6}\right)\rho^2\right]^{\frac{1}{2}}~.
\label{eq:jeans_star}
\eer
Note that $\mu, {\lambda_0 }$ and $m$ are not free parameters since they are related
through $m^2=\frac{2V_0\lambda^2}{m_{p}^2}$, ~ $\mu=\frac{3\lambda^3 V_0}{m_{p}^3}$,
${\lambda_0 }=\frac{7}{3}\frac{\lambda^4 V_0}{m_{p}^4}$.  The fundamental parameters in
this model are $V_0$ and $\lambda=\sqrt{\frac{2}{3\alpha}}$, with $\alpha$
defined in section \ref{sec:alphadm}.

\begin{figure}[H]
\centering
\scalebox{0.5}[0.48]{\includegraphics{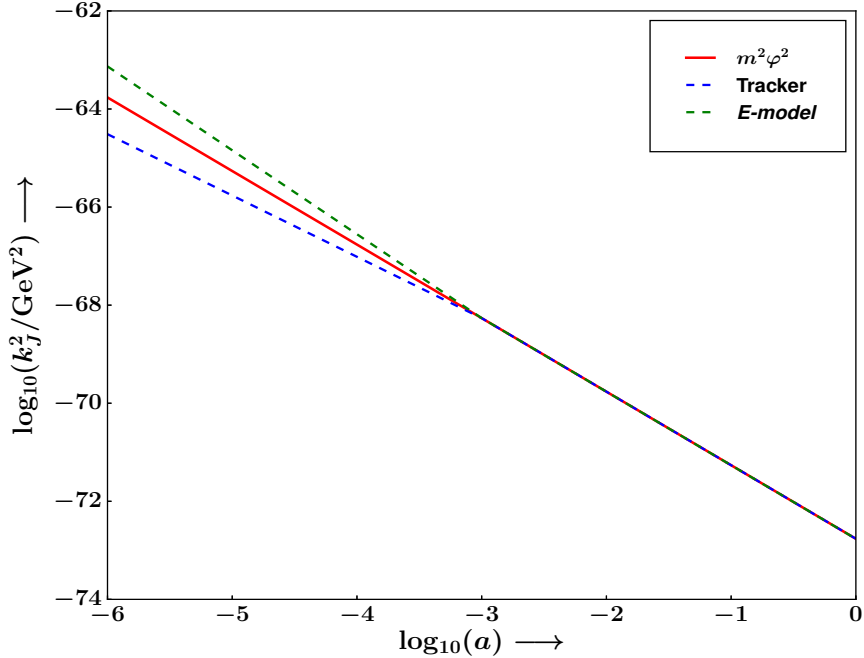}}
\caption{The evolution of the Jeans wavenumber $k_{J}^2$ is shown for the quadratic
potential (\ref{eq:jeans0}) with $m=10^{-22}\,{\rm eV}$ (red curve), the Tracker model
(\ref{eq:jeans3}) (blue curve) and the asymmetric {\em E-model\/} model (\ref{eq:jeans_star}) (green
curve). (In the Tracker and {\em E} models $\lambda=14.5$ which ensures
$m=10^{-22}\,{\rm eV}$. Note that, for a given value of $m$, only very fine-tuned
values of $\lambda$ ensure $\Omega_{0m} \sim 0.27$, as shown in figure~\ref{fig:Vmeff}.)
 We observe that $k_{J}^2$ converges to the same value in all three dark-matter models by
$z \sim 10^{3}$. For larger values of $m$, this convergence will occur at higher redshifts.
}
\label{fig:DMkJ}
\end{figure}

The expression for $k_{J}^2$ in each of the three $\alpha$-dark matter models
($\alpha$DM) considered by us, namely (\ref{eq:jeans_Tmodel}), (\ref{eq:jeans3}) and
(\ref{eq:jeans_star}), contains terms proportional to $\rho$ and $\rho^2$ (under a common
square root). By contrast, in the canonical $m^2\vphi^2$ model, $k_{J}^2$ is simply
proportional to $\sqrt{\rho}$, see (\ref{eq:jeans0}). Since $\rho^2$ falls off faster than $\rho$ during
expansion, it is important to study the evolution of $k_{J}^2$ in all of these dark
matter models. This has been done in figure~\ref{fig:DMkJ}, which shows $k_{J}^2$ for the
canonical $m^2\vphi^2$ model with $m=10^{-22}\,{\rm eV}$, the tracker model and the {\em E-model} (both with $\lambda=14.5$). This figure shows  that $k_{J}^2$ in all
three models converges to the  expression (\ref{eq:jeans0}) at late times by $z \sim 10^3$.

\begin{figure}[H]
\centering
\scalebox{0.48}[0.46]{\includegraphics{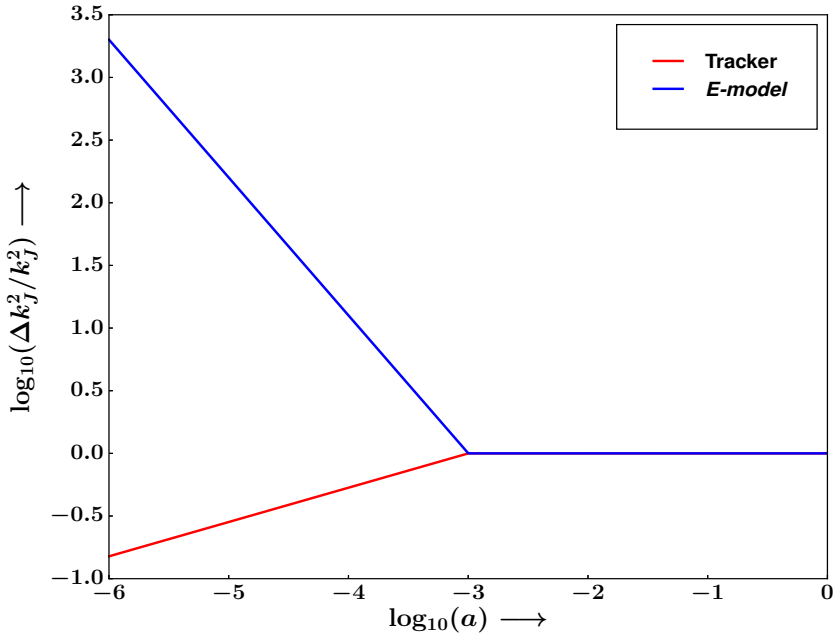}}
\caption{This figure shows the fractional difference between $k_{J}^2$ in the tracker
model (\ref{eq:jeans3}) (red curve) and the {\em E-model} (\ref{eq:jeans_star}) (blue
curve) on the one hand, and $k_{J}^2$ in the $m^2\vphi^2$ model (\ref{eq:jeans0})  on the
other ($m=10^{-22}\,{\rm eV}$ is assumed). Note that the fractional difference
${\Delta k_{J}^2}/{k_{J}^2}$ between the models becomes extremely small by redshift
$z\sim 10^3$, after which the clustering properties of the tracker and {\em E} models
become indistinguishable from those of the quadratic potential. For larger values of $m$,
the value of ${\Delta k_{J}^2}/{k_{J}^2}$ drops to zero at higher redshifts.}
\label{fig:DMdeltakJ}
\end{figure}

Another illustration of the result discussed above is depicted in figure~\ref{fig:DMdeltakJ}
which shows the fractional difference in $k_J^2$,
\ber
\frac{\Delta k_{J}^2}{k_{J}^2} = \frac{k_{J}^2\big|_{\alpha {\rm DM}} -
k_{J}^2\big\vert_{m^2\vphi^2}} {k_{J}^2\big\vert_{m^2\vphi^2}} \, ,
\eer
between the Tracker model and the {\em E-model} on the one hand, and the $m^2\vphi^2$
model on the other. Figures \ref{fig:DMkJ} and \ref{fig:DMdeltakJ} demonstrate that the
expression for the Jeans scale in the $m^2\vphi^2$ model, namely  (\ref{eq:jeans0}),
amply suffices to describe $k_J^2$ also in the Tracker and {\em E} models for $z <
10^3$.

\begin{figure}[H]
\centering
\scalebox{0.43}[0.41]{\includegraphics{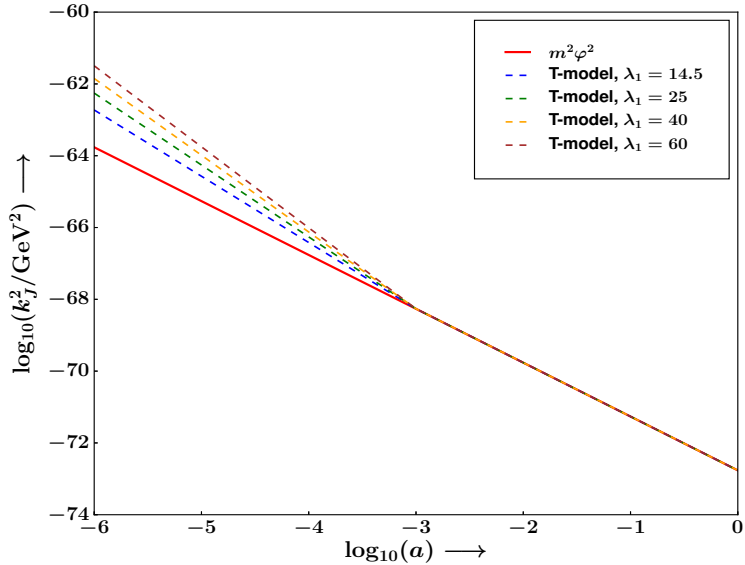}}
\caption{The Jeans wavenumber $k_{J}^2$ for the T-model (\ref{eq:jeans_Tmodel}) and the
quadratic potential (\ref{eq:jeans0}) is shown as a function of  $\log_{10} a$ for five
different values of $\lambda$. We find that, for all values of $\lambda_1$, the clustering
scale in the T-model converges to that in the model with $m^2\vphi^2$ potential by $z
\sim 10^3$. Note that $m=10^{-22}\,{\rm eV}$ in all models.} \label{fig:DMkJ1}
\end{figure}

Next we turn to the  T-model (\ref{eq:tanh}). In this case, one notes that there is no
bound on the values of $\lambda_1$ and $V_0$ as long as the mass constraint relation
(\ref{eq:starmass}) is satisfied. However, the expression for the Jeans scale in this
model, namely (\ref{eq:jeans_Tmodel}), was derived under the assumption that the harmonic
limit, $\lambda_1\vphi^2\ll m^2$, is valid. This helps place an upper limit on $\lambda_1$,
namely $\lambda<71$ if $m \simeq 10^{-22}\,{\rm eV}$. In figure~\ref{fig:DMkJ1}, the
Jeans scale $k_J^2$ for the T-model is shown as a function of the scale factor. This
figure shows that the expressions for $k_J^2$ in the T-model and $m^2\vphi^2$ model
converge for $z < 10^3$ (assuming identical values of $m$ in the two models). We
therefore conclude that, in all of the $\alpha$-dark matter models discussed by us, the
Jeans scale converges to that in the $m^2\vphi^2$ model, namely (\ref{eq:jeans0}), by $z
\sim 10^3$.

One also needs to draw attention to the following point. In all our dark matter models,
the scalar field begins to oscillate when $m/H \sim 1$. In other words, the scalar field
can commence oscillating (and become pressureless) once the expansion rate $H$ has
dropped below the scalar field mass $m$. Prior to this, the behaviour of $\vphi(t)$
depends upon the form of the potential. For the canonical potential $V =
\frac{1}{2}m^2\vphi^2$, the equation of state of the scalar field is $w_\vphi \simeq -1$
at early times. In
the case of tracker dark matter, $w_\vphi \simeq 1/3$ during the tracking phase, prior to
the onset of oscillations. %It is well known that, for a non-oscillatory scalar field, the
%speed of sound is the same as that of light. Consequently, it is unlikely that the
%dark-matter field $\vphi$ will cluster below the Hubble scale prior to oscillating. This
%will result in the introduction of a new length scale into the problem, in addition to
%the Jeans scale $k_J$ which comes into play later, during oscillations.
The EOS of matter during the pre-oscillatory epoch is likely to affect the power spectrum of dark matter perturbations. This is an important open problem  to which we wish to return in a companion paper.

%Its also worthwhile to draw attention to the following point. In all of the dark matter models
%alluded to above, the scalar field begins to oscillate when $\lambda\vphi \ll 1$.
%Let us denote the Hubble parameter when oscillations commence by $H_{\rm osc}$. Clearly at times
%earlier than $H_{\rm osc}^{-1}$ the scalar field will still be rolling down its (steep) potential.
% It is well
%known that for a non-oscillatory canonical scalar field
% the speed of sound is the same as that of light. %The Jeans length
%%then is of order the horizon size, consequently no the scalar field is stable to perturbations
%%on scales smaller than $H^{-1}$.
%This suggests that $\lambda_J^{osc}
%\sim c H_{\rm osc}^{-1}$ sets a length scale below which perturbations cannot grow.
%For the dark matter models discussed in this section,
%the larger of the two lengths, $k_J^{-1}$ and $c H_{\rm osc}^{-1}$, will therefore provide an estimate
%of the Jeans length.

\section{Dark Energy}
\label{sec:DE}

\subsection{Dark Energy from the potential $V \propto \left|\vphi\right|^{2n}$}
\label{sec:DEcano}

The canonical potential\footnote{Note that this potential also plays an important role in
monodromy inflation
\cite{monodromy}.} for oscillatory dark energy is given by  %(also see \cite{sami2002,nojiri2006})
\ber
V(\vphi)=V_0\left|\frac{\vphi}{m_p}\right|^{2n}\, , \qquad n<\frac{1}{2}~.
\label{eq:DEcano}
\eer
Once $\vphi$ commences oscillating around $\vphi = 0$, its time-averaged equation of
state becomes\footnote{DE in the context of $\alpha$-attractors has also been investigated in \cite{linder,Kostas_owen}. In \cite{Kostas_owen} $\alpha$-attractors were used to construct a model of Quintessential-Inflation, while Quintessential-Inflation in the context of oscillating DE has been discussed in \cite{sami2002,nojiri2006}.}
\ber
\langle w_\vphi\rangle = \frac{n-1}{n+1}~.
\label{eq:DEEOS}
\eer
The oscillating scalar field can drive cosmic acceleration when $\langle w_\vphi\rangle <
-1/3$, which corresponds to $n < {1}/{2}$.
However, as we shall soon show, the %just as in the case of dark matter ($n=1$), the
initial density of the scalar field requires a high degree of fine-tuning in order to account for the
current value of dark energy parameter $\Omega_{0,{\rm DE}} \sim 0.7$. %(Other aspects of oscillating DE models have been discussed in \cite{sami2002,nojiri2006}.)

\begin{figure}[H]
\centering
\includegraphics[width=0.75\textwidth]{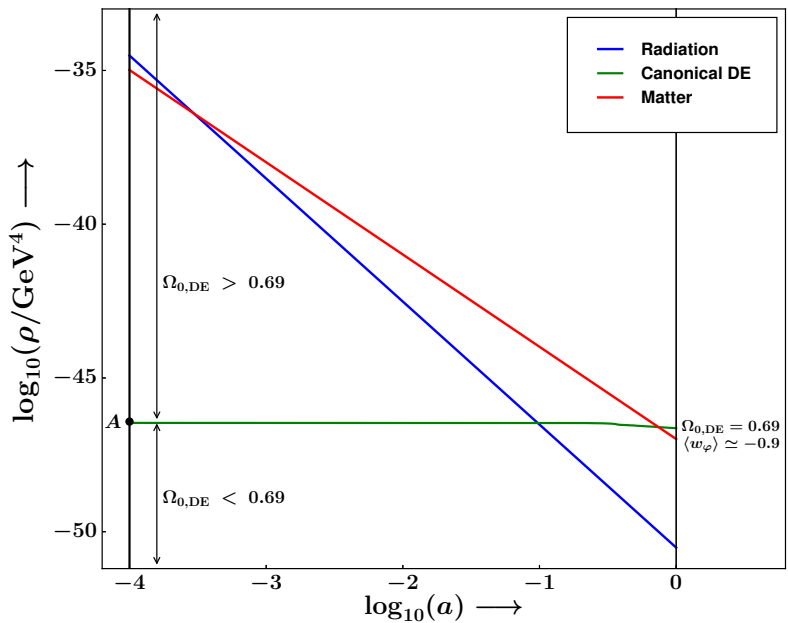}
\caption{This figure describes the fine tuning of the initial scalar-field energy density
associated with dark energy. Commencing our integration at $a=10^{-4}$ ($z = 10^{4}$), we
find that the scalar-field energy density remains frozen to its initial value all the way
until $z \sim 2$. Thereafter, due to the rapid decline in the matter density, the damping
on the scalar field gets lifted and $\vphi$  begins to oscillate and behave like dark
energy with $\langle w_\vphi\rangle \simeq -0.9$. The initial scalar-field energy density
which results in $\Omega_{0,{\rm DE}} \simeq 0.69$ is shown by the  point $A$. Initial
values of $\rho_i < \rho_A$ lead to too little dark energy at the present epoch
($\Omega_{0,{\rm DE}} < 0.69$), whereas $\rho_i > \rho_{A}$ leads to too much dark energy
($\Omega_{0,{\rm DE}} > 0.69$). } \label{fig:DEcan1}
\end{figure}

% We assume the conservative value  $n=0.01$ for which $\langle w\rangle\simeq -0.98$, which is
%in excellent agreement with current observational constraints \cite{Planck_DE}.
The following point deserves mention in a discussion of DE sourced by an oscillating
scalar field. It has been noted that such models are prone to developing a dynamical
instability  \cite{kamion}. Therefore, in order to remain viable, the scalar field in
these models must begin oscillating at late times, $z_{\rm osc}<5$. This ensures that the
field oscillates only  a few times before the present epoch thereby limiting the growth
of the instability \cite{dutta_scherrer}. However, the extreme case when oscillations
commence very late, $z_{\rm osc}<1$, leads to a situation in which the phase of
oscillations begins to play an important role and the time-averaged formula
(\ref{eq:DEEOS}) is no longer valid \cite{dutta_scherrer}. Taking account of these
different factors, we have chosen, for $n=0.05$, $V_0\sim 5\times 10^{-47}\,{\rm GeV}^4$,
thereby ensuring that oscillations commence by $z_{\rm osc}\sim 2$. Our results,
summarized in figure~\ref{fig:DEcan1} demonstrate that the (fine-tuned) initial value
$\rho_\vphi$ which results in $\Omega_{0, {\rm DE}}=0.69$, is given by
\ber
 \rho_A=4.51\times 10^{-47}\,{\rm GeV}^4~.
\eer

 %Evolution of denisty of the scalar field with the scale factor has been described in figure \ref{fig:DEstarL} by the black color dotted and dashed line where point A represents the fine tunned initial value of denisty.

We therefore find that the canonical potential (\ref{eq:DEcano}) suffers from a severe
fine-tuning problem since, for a given value of $n$ and $V_0$, there is only a
very narrow range of values\footnote{Note that, as in the case of canonical dark
matter in section \ref{sec:canoDM}, the fine tuning which we observe is insensitive to
the initial value of $\dot{\vphi}$. } of the initial energy density which lead to the
current value of dark energy density $\Omega_{0,{\rm DE}}$. This is very similar to the
fine tuning which afflicts dark matter with the $m^2\vphi^2$ potential,
 discussed in section \ref{sec:canoDM}.

\begin{figure}[H]
%\centering
\hspace{-1.2cm}
%\subfloat[ ][ ]
{\includegraphics[width=0.55\textwidth]{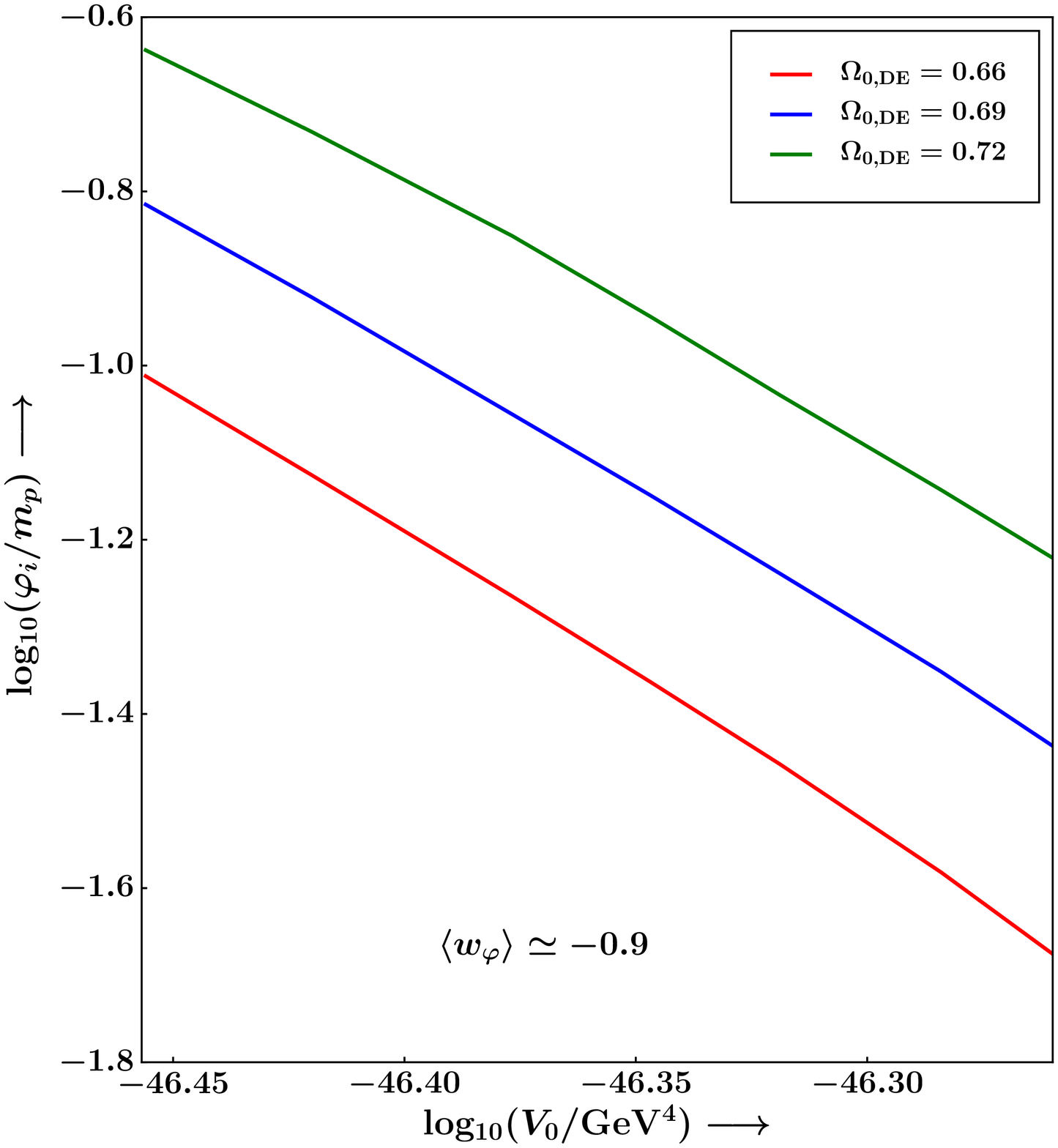}}
%\subfloat[ ][ ]
{\includegraphics[width=0.55\textwidth]{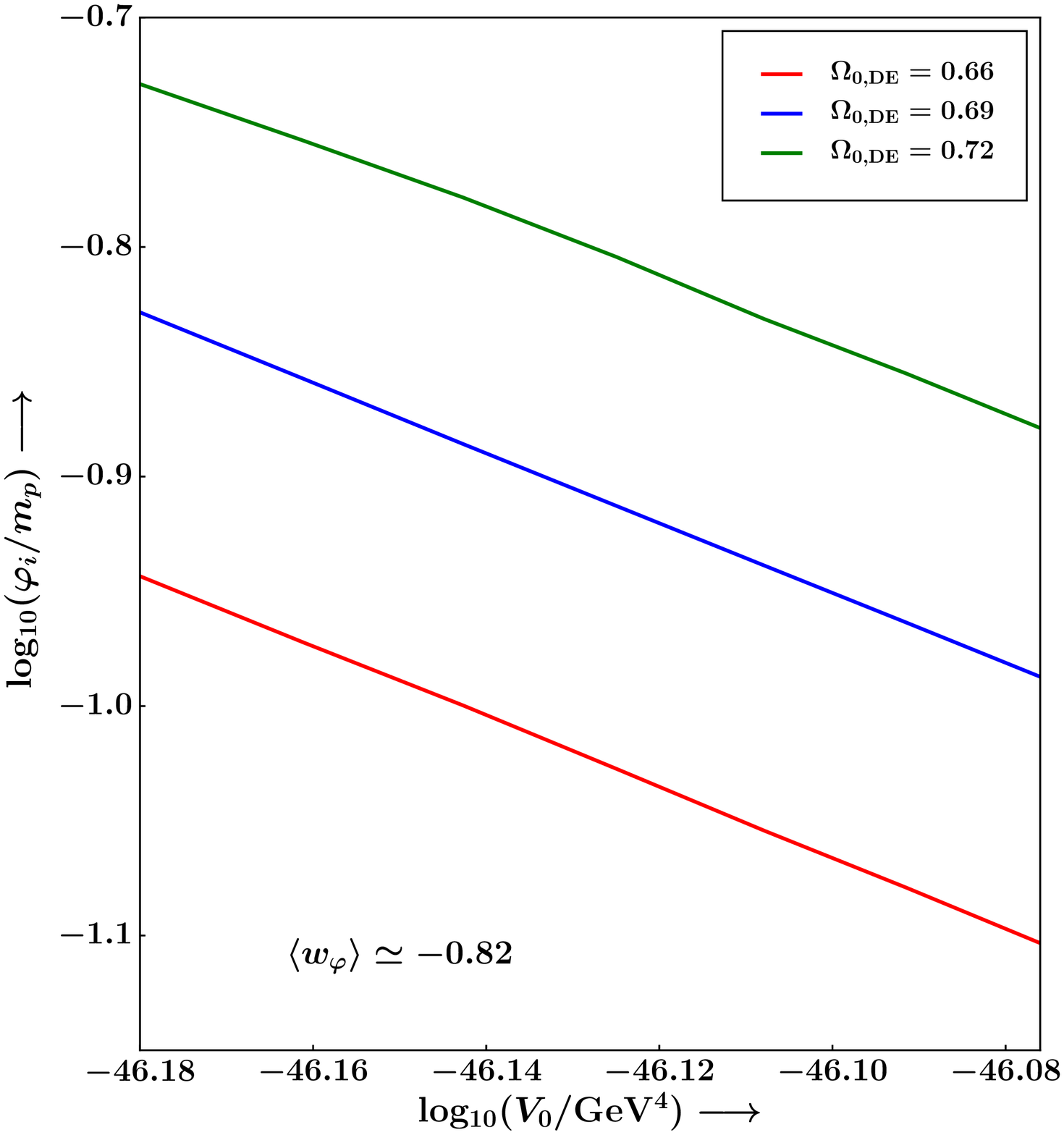}}
\caption{Dependence of the initial value of the scalar field $\vphi_i$ on  $V_0$ is shown
for the canonical dark-energy potential
$V(\vphi)=V_0\left|{\vphi}/{m_p}\right|^{2n}$. The left panel ($a$) and  the right
panel ($b$) correspond to $n=0.05$ and $0.1$, respectively. Three different values of the
present dark energy density $\Omega_{0,{\rm DE}}$ are assumed. One sees that, for a given
value of $\Omega_{0,{\rm DE}}$, the value of $\vphi_i$ decreases with increasing $V_0$.
Whereas for a fixed value of  $V_0$, larger values of $\Omega_{0,{\rm DE}}$ are
associated with larger initial values of $\vphi_{i}$. } \label{fig:DEcanoPM}
\end{figure}

Next, we determine the effect of varying $V_0$ on the initial
field value $\vphi_i$. Figure~\ref{fig:DEcanoPM} illustrates the relationship between
$\vphi_i$ and $V_0$ for two different cases: (i) $n=0.05 \Rightarrow \langle
w_\vphi\rangle \simeq -0.9$, (ii) $n=0.1 \Rightarrow \langle w_\vphi\rangle \simeq
-0.82$. We observe that larger values of $\Omega_{0,{\rm DE}}$ requires a larger
$\vphi_i$ for a given $V_0$. (Note that ${\dot\vphi} \simeq 0$ initially, since the
motion of the dark-energy field is heavily damped, initially by radiation and then by
matter. As a result, $\rho_i \simeq V_0 \left\vert {\vphi_i}/{m_p} \right\vert^{2n}$.)

 \subsection{Dark Energy from the asymmetric {\em E-Model}}
 \label{sec:DEstar}

\begin{figure}[H]
\centering
\includegraphics[width=0.68\textwidth]{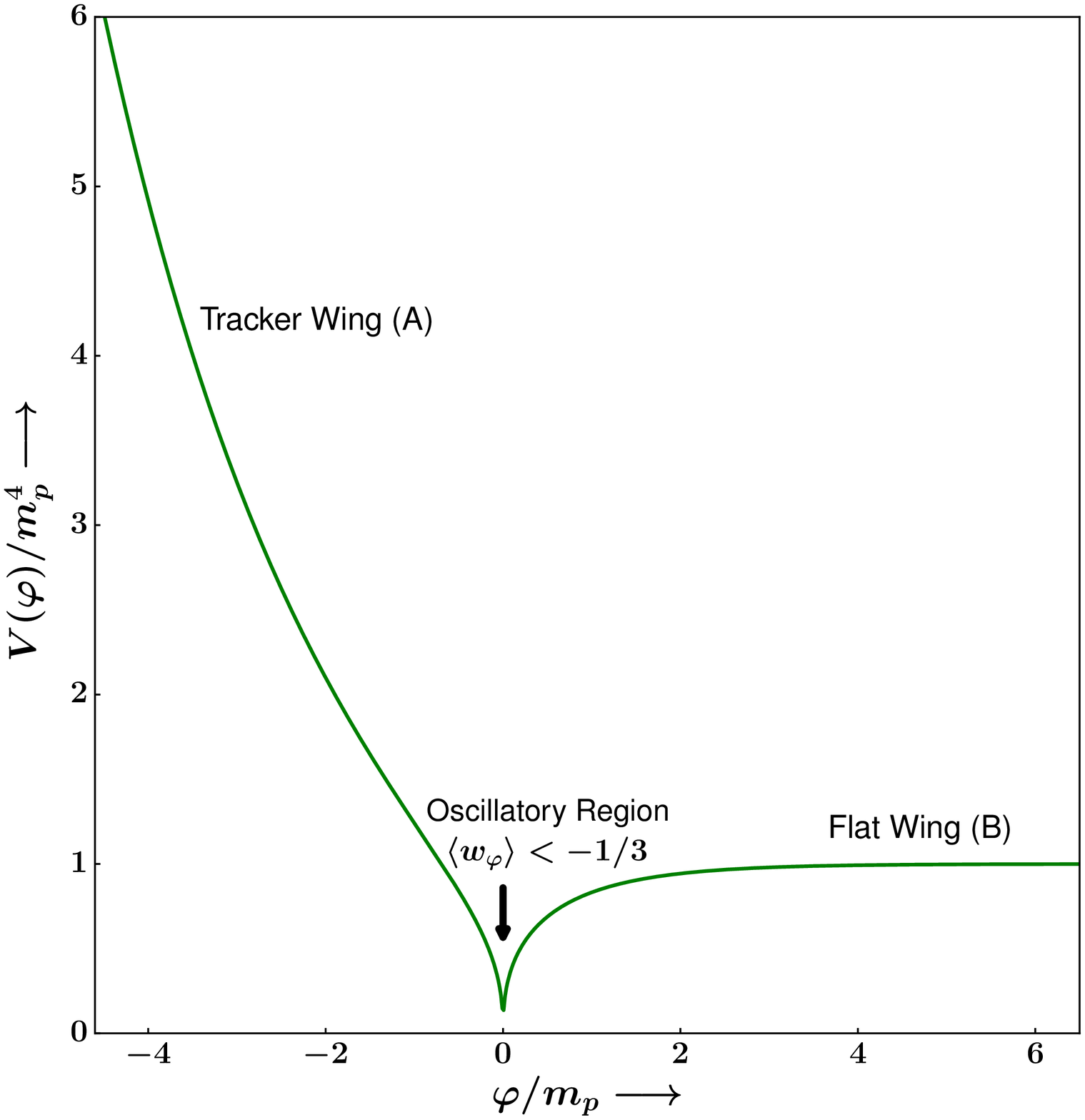}
\caption{This figure schematically illustrates the {\em E-model} potential
(\ref{eq:DEstar}) with $\lambda=1$ and $n=0.2$. The main features of this potential are:
(A) the exponential tracker wing for $\lambda|\vphi| \gg m_p$ ($\vphi<0$), (B) the flat
wing for $\lambda\vphi \gg m_p$, and the oscillatory region for which $\lambda|\vphi| \ll
m_p$, so that $V(\vphi) \simeq V_0\left|{\vphi}/{m_p}\right|^{2n}$. }
\label{fig:starpotDE}
\end{figure}

As we saw earlier, the {\em E-model} has features common to both the T-model and
tracker DE\@. We shall therefore confine our attention solely to this model in our
investigation of DE from $\alpha$-attractors since its properties will carry over to the
other two models as well. The potential
\ber
V(\vphi)=V_0\left(1-e^{-\lambda\frac{\vphi}{m_p}}\right)^{2n}\, , \qquad n<\frac{1}{2} \,
, \label{eq:DEstar}
\eer
exhibits three asymptotic regions given by (see figure~\ref{fig:starpotDE})
\ber
\mbox{tracker wing:} \quad V(\vphi) &\simeq& V_0\, e^{2n\lambda|\vphi|/m_p}~, \quad \vphi
< 0, \quad \lambda|\vphi| \gg m_p \, ,
\label{eq:DEstarpot1}\\
\mbox{flat wing:} \quad V(\vphi) &\simeq& V_0\, , \quad \lambda\vphi \gg m_p \, ,
\label{eq:DEstarpot2}\\
\mbox{oscillatory region:} \quad V(\vphi) &\simeq&
V_0\left|\frac{\vphi}{m_p}\right|^{2n}\, , \quad \lambda|\vphi| \ll m_p \, .
\label{eq:DEstarpot3}
\eer
The relation between $\lambda$ and $\alpha$ is given in (\ref{eq:lambda-alpha}).
We focus first on the tracker wing of the potential, %(flat wing is not very appealing from the point of view of generic initial conditions as described in the case of $\alpha$-DM) which corresponds to
for which
\ber
\frac{\rho_{\vphi}}{\rho_{\rm total}}=\frac{1}{n^2\lambda^2} \quad &&\mbox{during
radiation domination} \, , \label{eq:DEtrackR}\\
\frac{\rho_{\vphi}}{\rho_{\rm total}}=\frac{3}{4n^2\lambda^2} \quad &&\mbox{during matter
domination} \, . \label{eq:DEtrackM}
\eer

\begin{figure}[H]
%\centering
\hspace{-2cm}
%\subfloat[ ][ ]
{\includegraphics[width=0.65\textwidth]{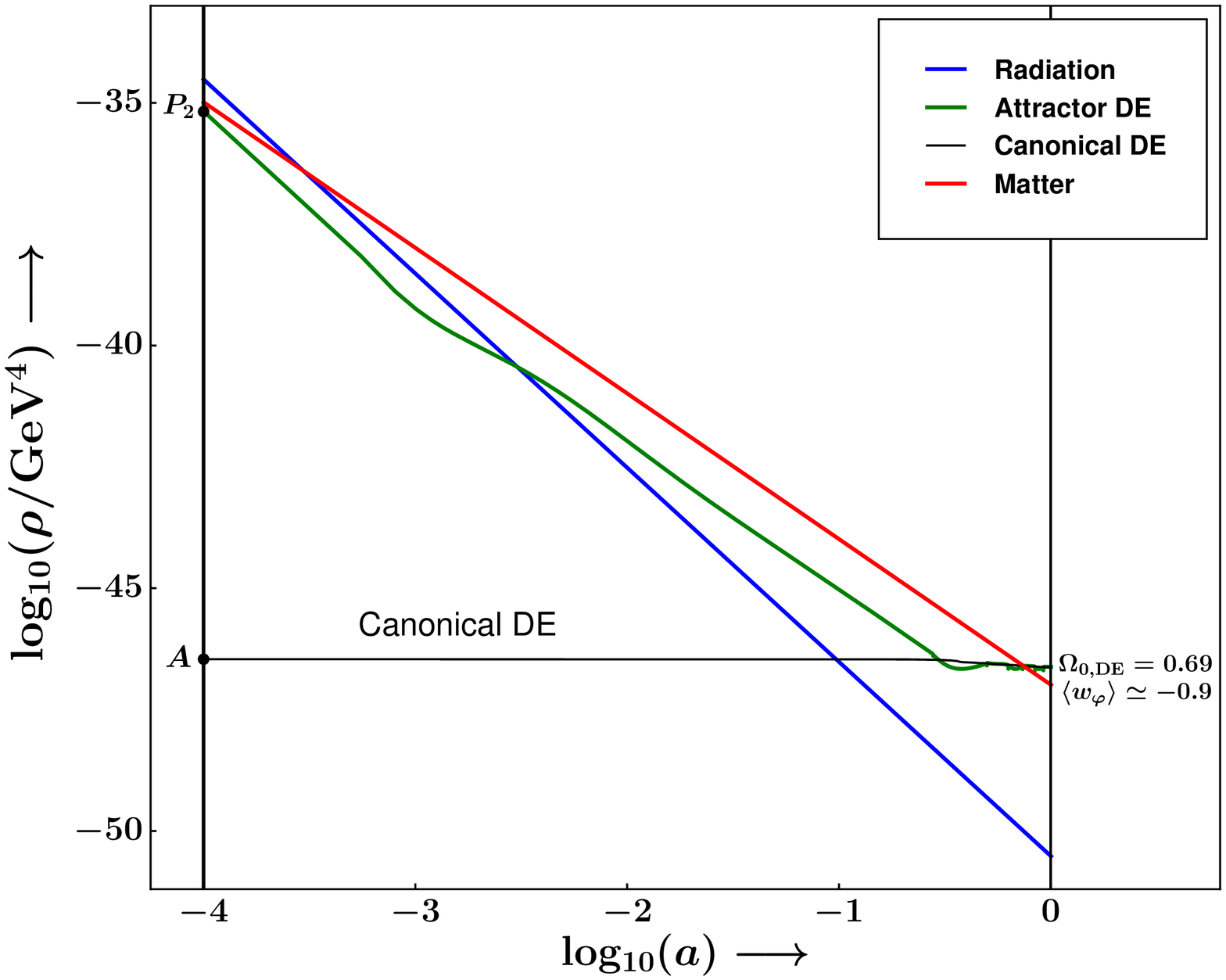}}
%\hspace{-0.05cm}
%\subfloat[ ][ ]
{\includegraphics[width=0.645\textwidth]{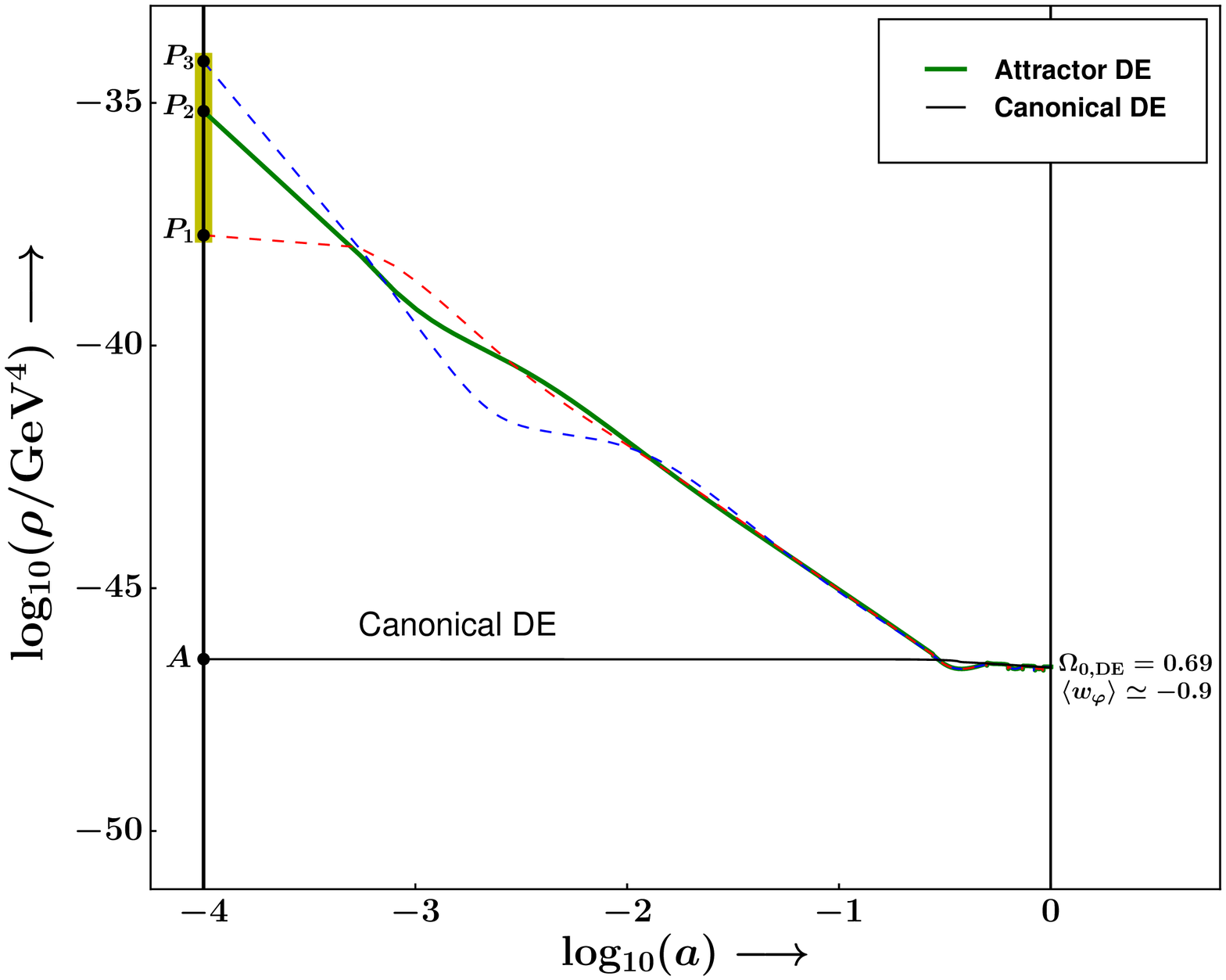}}
\caption{This figure describes the evolution of the energy density in radiation, matter
and a scalar field, with the latter playing the role of DE\@. The scalar field commences
its descent from the tracker wing  of the potential $V(\vphi) = V_0 \left ( 1 -
e^{-\lambda\vphi/m_p}\right )^{2n}$ with $n=0.05$, $\lambda=60$ and $V_0=2.56\times
10^{-47}\,{\rm GeV}^4$. These parameter values ensure $\langle w_{\vphi}\rangle \simeq
-0.9$ and $\Omega_{0,{\rm DE}} = 0.69$. {\em Left panel (a):} The green line represents
 DE which tracks the radiation and matter energy densities during the radiative
and matter-dominated epochs, respectively. The near-horizontal black line corresponds to
canonical DE (\ref{eq:DEcano}) whose fine-tuned initial value is represented by point $A$. {\em Right
panel:} The green band spanning from $P_{1}$ to $P_{3}$ shows  the range in initial values of
the scalar-field  density (at $z=10^{4}$) that lead to the current value
$\Omega_{0,{\rm DE}}=0.69$. $P_{2}$ marks the initial  density
corresponding to the attractor solution (green line) towards which all trajectories
commencing from the $P_1$--$P_3$ band converge. This scenario is in sharp contrast to the
case of the canonical model with $V(\vphi)=V_0\left|{\vphi}/{m_p}\right|^{2n}$, shown in
figure~\ref{fig:DEcan1}, for which only finely tuned initial values of the
energy density (point $A$),  lead to $\Omega_{0,{\rm DE}} \simeq 0.69$. (For clarity of
presentation, we do not show the baryon density in either panel.) } \label{fig:DEstarL}
\end{figure}

As discussed earlier, in order to evade the instabilities plaguing a cuspy potential such
as (\ref{eq:DEstarpot3}), the field must commence oscillating late, by $z_{\rm osc} < 5$.
Choosing $n=0.05 \Rightarrow \langle w_\vphi\rangle \simeq -0.9$, the values of $V_0$ and
$\lambda$ which satisfy this constraint and which also yield $\Omega_{0,{\rm DE}}=0.69$
are $V_0=2.56\times 10^{-47}\,{\rm GeV}^4$, $\lambda=60$ (hence $\alpha=1.85\times
10^{-4}$). As in the case of $\alpha$DM, we find that DE based on the {\em E-model}
converges to the late-time attractor $V(\vphi) \sim \vphi^{2n}$ from a wide range of
initial conditions. This is illustrated in figure~\ref{fig:DEstarL}. Commencing at
$a=10^{-4}$ ($z \sim 10^4$), we find that  initial energy-density values spanning almost
4 orders of magnitude converges to the attractor curve (green line) yielding $\Omega_{0,
{\rm DE}}=0.69$; see the right panel of figure~\ref{fig:DEstarL}.

The attractor value of the initial density (marked by point $P_2$ in
figure~\ref{fig:DEstarL}) is given by $\rho_{P_2}=6.82\times 10^{-36}\,{\rm GeV}^4$,
while the maximum and minimum values of the initial density which yield  $\Omega_{0, {\rm
DE}}=0.69$ are
\ber
\rho_{\rm max} &=& \rho_{P_3}=3.98\times 10^{-35}~{\rm GeV}^4 \, , \\
\rho_{\rm min} &=& \rho_{P_1}=1.59\times 10^{-38}~{\rm GeV}^4~.
\eer
The range in the attractor $P_1$--$P_3$ band will clearly increase if we commence evolving our system of equations from a
more realistic  earlier epoch. For instance, the range in the energy-density values
prescribed at the GUT scale ($z\sim 10^{26}$) which converges to $\Omega_{0, {\rm
DE}}=0.69$ covers more than $112$ orders of magnitude. The fact that such an impressively
large range of initial conditions  can lead to the present universe removes the extreme
fine-tuning associated with the potential (\ref{eq:DEcano}) and makes DE from tracker
potentials more compelling.

\begin{figure}[H]
\centering
\includegraphics[width=0.66\textwidth]{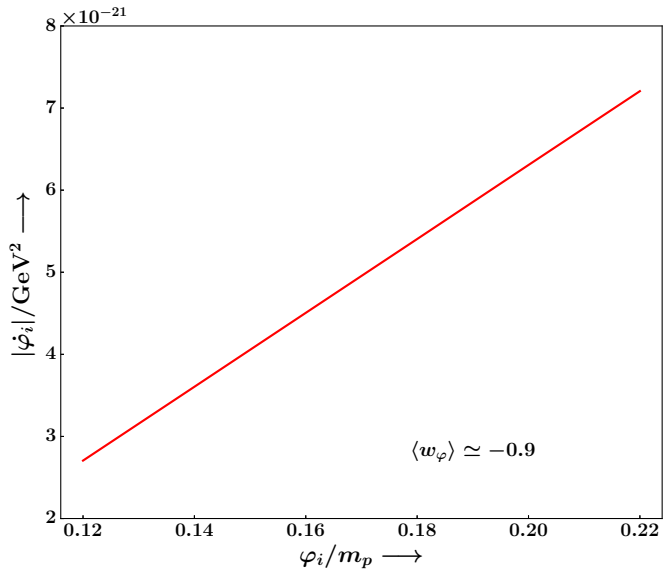}
\caption{The initial velocity $\dot{\vphi_i}$ which results in $\Omega_{0,{\rm DE}}=0.69$
is plotted as a function of $\vphi_i$ for the flat wing of the asymmetric {\em E-model} potential
(\ref{eq:DEstar}). The parameter values are $V_0=2.56\times 10^{-47}\,{\rm GeV}^4$,
$\lambda=60$ and $n=0.05 \Rightarrow \langle w_\vphi\rangle \simeq -0.9$. Notice that  a
larger initial value of the scalar field $\vphi_i$ requires a larger value of
$|\dot{\vphi_i}|$ to ensure $\Omega_{0,{\rm DE}}=0.69$. } \label{fig:DEstarflat}
\end{figure}

Turning to the flat right wing of the potential (\ref{eq:DEstar}), one notices that the
initial velocity $\dot{\varphi_i}$ plays an important role in the analysis of initial
conditions just as it did for dark matter (see section \ref{sec:flat}). Commencing on the
flat wing, a given initial value $\vphi_i$ requires a particular value of
$\dot{\varphi_i}$ in order to give rise to $\Omega_{0,{\rm DE}}=0.69$. The dependence of
the initial velocity $\dot{\varphi_i}$ on $\vphi_i$ is shown in
figure~\ref{fig:DEstarflat}. This figure shows that that a larger $\vphi_i$, associated
with a point further along the flat wing, requires a larger value of $|\dot{\vphi_i}|$ in
order to arrive at a given $\Omega_{0,{\rm DE}}$ (also see figure~\ref{fig:flat4}).

We therefore conclude that, from the viewpoint of initial conditions, the flat wing of the {\em E-model} potential
results in a  less appealing model of DE as compared to the steep wing, for which one arrives at
a given $\Omega_{0,{\rm DE}}$ from a much wider class of initial conditions.

\section{Discussion}
\label{sec:summary}

%\begin{itemize}

In this paper, we have demonstrated that the $\alpha$-attractors, which provide
interesting models of inflation \cite{linde1,linde2}, can also lead to compelling models
of dark matter and perhaps even dark energy. It is well known that a simple scalar field
model of dark matter can be constructed from the potential usually associated with
chaotic inflation, namely $V = \frac{1}{2}m^2\vphi^2$. For an ultra-light mass of $m \sim
10^{-22}\,{\rm eV}$, this model can successfully resolve several key problems faced by
the standard cold dark matter scenario including the cusp--core issue, the problem of
substructure etc.\@ \cite{sahni_wang,hu00,Witten}. However, what is sometimes overlooked
is that this interesting model also suffers from a severe fine-tuning issue. As
demonstrated in section \ref{sec:canoDM}, for a given value of $m$, the
region of initial values of $\vphi_{i}$ which
leads to the current value of the matter density $\Omega_{0m}$ is \underline{exceedingly small}. Moreover the fine tuning discussed here is expected to carry over to other potentials with an $m^2\vphi^2$ asymptote including axionic dark matter.

This difficulty is avoided if dark matter is sourced by $\alpha$-attractors. Of the
$\alpha$-attractor models studied in sec. \ref{sec:dm}, the {\em E-model\/}
(\ref{eq:star}) and the tracker potential (\ref{eq:sw}) provide the
%tracker branch of the
%Starobinsky potential (\ref{eq:star}) and the tracker potential (\ref{eq:sw}) provide the
most versatile examples of scalar-field dark matter. In both, one arrives at the
late-time dark matter asymptote from a very wide range of initial conditions. An
important feature of the asymmetric {\em E-model\/} is the presence of a steep
non-inflationary wing of the potential in addition to the inflationary plateau,
 as shown in figure \ref{fig:staroughpot}. The required steepness of the tracker wing is
supported by a small value of the geometric parameter $\alpha \leq 10^{-3}$. This is the
main difference with the Starobinsky model $R+M^2R^2$, which corresponds to $\alpha=1$.
An analogous requirement is valid for the tracker potential (\ref{eq:sw}). These two
$\alpha$-attractor models, therefore, successfully overcome the fine-tuning problems
associated with the $m^2\vphi^2$ potential.

Another important feature of $\alpha$-dark matter, shared by the
$m^2\vphi^2$ potential, is that the Jeans length associated with gravitational clustering
can be quite large.  Our analysis in section \ref{sec:grav} demonstrates  that, despite significant differences in dynamics, the Jeans scale in all of our DM models converges to the same late-time expression (\ref{eq:jeans0}). We therefore conclude that {\em tracker based\/}
$\alpha$-dark matter shares all of the distinctive features of {\em fuzzy\/} dark matter
(sourced by the $m^2\vphi^2$ potential), without having any of its limitations.

Finally, we would like to caution the reader that our attempt in this paper has not been
to create a unified description of inflation and dark matter based on
$\alpha$-attractors. It has been more modest. We have simply shown that the very same
potentials which characterize $\alpha$-attractors can also be used to construct
compelling models of dark matter, and perhaps even dark energy. Nevertheless, one can, in
principle, envisage the possibility that dark matter arose from the coherent oscillations
in the $\alpha$-attractor inflaton after reheating. This follows from the observation
that all of the $\alpha$-attractor potentials have a minimum near which $V \simeq
\frac{1}{2}m^2\vphi^2$, which permits the inflaton to oscillate. In the standard
scenario, the universe preheats during this oscillatory phase due to the coupling of $\vphi$
to the standard
model fields. %The inflaton commeces oscillating when $\vphi$ drops below $m_{\rm Pl}$ and
%$m>H$.
Preheating transfers the energy of the inflaton into new degrees of freedom
(quarks and leptons, radiation etc.\@)  created during preheating. % and which give
%rise to the universe we live in.
Consequently, the value of the inflaton drops drastically from $\vphi \sim m_p$ at the
commencement of preheating to $\vphi_{\rm end} \ll m_p$ at its end. If $\vphi_{\rm end}
\neq 0$ (i.e., if preheating is not complete), then the post-preheating
oscillations of $\vphi$ will cause the inflaton density to drop as $\rho_\vphi \propto
a^{-3}$ a consequence of the fact that $\langle w_\vphi\rangle \simeq 0$. In other words,
the inflaton can now play the role of dark matter \cite{preheating,preheating1}. It is
straightforward to show that if the universe preheats to a temperature $T_r$ then, at
$z_r = T_r/T_0 - 1$, the dark-matter field must satisfy the relation
\ber
\frac{1}{2}m^2\vphi^2_r \simeq \rho_r(z_r) \frac{T_{0\gamma}}{T_\gamma(z_r)}\frac{\Omega_{0m}}{\Omega_{0r}},
\eer
where $\vphi_r \equiv \vphi(z_r)$ and $\rho_r = \sigma T_r^4$ is the radiation density immediately after preheating.
In other words
\ber
\vphi_r \simeq \left\lbrack
\frac{2\rho_r(z_r)}{m^2}\frac{T_{0\gamma}}{T_\gamma(z_r)}\frac{\Omega_{0m}}{\Omega_{0r}}
\right\rbrack^{1/2}~.
\eer
Post-preheating
scalar field values $\vphi > \vphi_r$ will lead to an excess of dark matter,
whereas $\vphi < \vphi_r$ will result in too little dark matter at the current epoch.
A similar fine tuning of initial conditions for the $m^2\vphi^2$ potential was noted earlier in
section \ref{sec:canoDM}.

The reader might also note that our attempt in this paper has been to investigate the many
different features associated with $\alpha$DM and $\alpha$DE\@. We have intentionally
refrained from testing these models against observations, leaving this exercise to a
future paper.

%\end{itemize}

%\ber
%{\cal L} = \sqrt{-g}\left\lbrace \frac{1}{2}\partial_\mu\chi\partial^\mu\chi
%+ \frac{\chi^2}{12}R(g) - \frac{1}{2}\partial_\mu\phi\partial^\mu\phi
%-\frac{\phi^2}{12}R(g) - \frac{\lambda}{4}\left (\phi^2 - \chi^2\right )^2\right\rbrace
%\label{eq:1}
%\eer

\section*{Acknowledgments}
S.S.M. thanks the Council of Scientific and Industrial Research (CSIR), India, for
financial support as senior research fellow.

\appendix

\section{Linear perturbations and instability}
\label{app:instab}

The Friedmann--Robertson--Walker metric with scalar perturbations has the form
\ber
ds^2 = (1 + 2\Psi) d t^2 - a^2 (t) (1 - 2 \Psi) d {\vec x}^{\,2} \, .
\eer
The linearized equations for perturbations read
\ber \label{pert1}
&&\delta \ddot \varphi + 3 H \delta \dot \varphi + V'' (\varphi) \delta \varphi -
\frac{\nabla^2}{a^2} \delta \varphi = 4 \dot \varphi \dot \Psi - 2 V' (\varphi) \Psi \, , \\
&&\frac{\nabla^2}{a^2} \Psi = \frac{1}{2 m_{p}^2} \left[ \dot \varphi \delta \dot \varphi +
V' (\varphi) \delta \varphi + 3 H \dot \varphi \delta \varphi - \dot \varphi^2 \Psi
\right] \, , \label{pert2}
\eer
where an overdot denotes the derivative with respect to the time $t$.  Here, we use the
background equations
\ber
H^2 = \frac{\rho}{3 m_{p}^2} = \frac{1}{3 m_{p}^2} \left[ \frac12 \dot \varphi^2 + V
(\varphi)\right] \, , \qquad \dot H = - \frac{(\rho + p)}{2 m_{p}^2} = - \frac{\dot
\varphi^2}{2 m_{p}^2} \, ,
\eer
and
\ber
\ddot \varphi + 3 H \dot \varphi + V' (\varphi) = 0 \, ,
\eer
with $H = \dot a / a$.  All equations are also valid  if the potential $V (\varphi)$
includes the cosmological constant.

In the domain of interest,
\ber \label{domain}
H^2, |\dot H| \ll \frac{k^2}{a^2} \ll m^2 \, ,
\eer
where $m$ is the mass of the scalar field, several terms in the system of equations (\ref{pert1}),
(\ref{pert2}) can be neglected so that this system simplifies to
\ber \label{pert11}
&&\delta \ddot \varphi + V'' (\varphi) \delta \varphi - \frac{\nabla^2}{a^2} \delta
\varphi \approx 4 \dot \varphi \dot \Psi - 2 V' (\varphi) \Psi \, , \\
&&\frac{\nabla^2}{a^2} \Psi \approx \frac{1}{2 m_{p}^2} \left[ \dot \varphi \delta \dot
\varphi + V' (\varphi) \delta \varphi \right] \, . \label{pert12}
\eer
Proceeding to the Fourier transform and systematically using (\ref{domain}), we have from
(\ref{pert12}):
\ber
\Psi \approx - \frac{a^2}{2 m_{p}^2 k^2} \left[ \dot \varphi \delta \dot \varphi + V'
(\varphi) \delta \varphi \right] \, , \qquad \dot \Psi \approx - \frac{a^2 \dot
\varphi}{2 m_{p}^2 k^2} \left[ \delta \ddot \varphi + V'' (\varphi) \delta \varphi \right] \,
.
\eer
Substituting this into (\ref{pert11}), we obtain a closed approximate equation for
$\delta \varphi$\,:
\ber \label{closed}
\delta \ddot \varphi + V'' (\varphi) \delta \varphi + \frac{k^2}{a^2} \delta \varphi -
\frac{a^2 V' (\varphi)}{m_{p}^2 k^2} \left[ \dot \varphi \delta \dot \varphi + V' (\varphi)
\delta \varphi \right] = 0 \, .
\eer

One can see that the approximation (\ref{domain}) practically means that we can neglect
the cosmological expansion altogether for the wavenumbers $k$ of interest and also that
we use formal perturbation theory to the leading order of the gravitational coupling $1 /
m_{p}^2$.  The change of variables
\ber
\delta \varphi = \left[ 1 + \frac{a^2 V (\varphi)}{2 m_{p}^2 k^2} \right] \upsilon
\eer
in (\ref{closed}) eliminates, in the same approximation, the term with the first derivative
for $\upsilon$, and we obtain
\ber \label{veq}
\ddot \upsilon + \left[ V'' +  \frac{k^2}{a^2} - \frac{a^2}{2 m_{p}^2 k^2} \left( 3 V'{}^2 -
V'' \dot \varphi^2 \right) \right] \upsilon = 0 \, .
\eer

In connection with our approximation, we set $a = 1$ and investigate this equation for
different potentials $V (\varphi)$.

$\bullet$ In the simplest case
\ber
V (\varphi) = \frac12 m^2 \varphi^2\, ,
\eer
the scalar field oscillates as $\varphi = \varphi_0 \cos m t$, and we get the equation
\ber
\ddot \upsilon + \left( m^2 +  k^2 - \frac{m^4 \varphi_0^2}{2 m_{p}^2 k^2} - \frac{m^4
\varphi_0^2}{m_{p}^2 k^2} \cos 2 m t \right) \upsilon = 0 \, .
\eer
By proceeding to the dimensionless time variable $\tau = m t$, we transform this into the
canonical Mathieu equation
\ber \label{Mathieu}
\ddot \upsilon + \left( P - 2 Q \cos 2 \tau \right) \upsilon = 0
\eer
with
\ber
P = 1 + \frac{k^2}{m^2} - \frac{m^2 \varphi_0^2}{2 m_{p}^2 k^2} \, , \qquad Q = \frac{m^2
\varphi_0^2}{2 m_{p}^2 k^2} \, .
\eer
Our domain of values of $k$ ensures that the value of $P$ lies very close to unity.  In
this case, the instability domain for the Mathieu equation corresponds to
\ber \label{instab}
|Q| > |P - 1| \, .
\eer
We obtain the condition which coincides with the result of \cite{hu00}:
\ber
k^4 < \frac{m^4 \varphi_0^2}{m_{p}^2} = \frac{2 m^2 \rho_\varphi}{m_{p}^2} \, .
\eer

$\bullet$ In the case
\ber
V (\varphi) = \frac12 m^2 \varphi^2 + \frac14 \lambda_0 \varphi^4
\eer
with the additional condition $\lambda_0 \varphi^2 \ll m^2$, we can include the
corresponding correction to the leading part in (\ref{veq}), that survives in the limit
$m_{p} \to \infty$. This results in the same Mathieu equation (\ref{Mathieu}) with $P$
and $Q$ shifted with respect to the preceding case as $P \to P + 3 \lambda_0 \varphi_0^2 /
2m^2$, $Q \to Q - 3 \lambda_0 \varphi_0^2 / 4m^2$.  Condition (\ref{instab}) then reads
\ber \label{quartic}
k^4 + \frac94 \lambda_0 \varphi_0^2 k^2 < \frac{m^4 \varphi_0^2}{m_{p}^2} \, ,
\eer
which, apart from the coefficient of the second term on the left-hand side, essentially
coincides with the result of \cite{kamion}.

$\bullet$ Finally, consider the case with a cubic term in the potential, i.e.,
\ber
V (\varphi) = \frac12 m^2 \varphi^2 - \frac13 \mu \varphi^3 + \frac14 \lambda_0 \varphi^4
\eer
with the condition $\lambda_0 \varphi^2 , \mu \varphi \ll m^2$.  In this case, the main
equation (\ref{veq}) reads
\ber \label{veq1}
\ddot \upsilon + \left[ m^2 +  k^2 + \frac32 \lambda_0 \varphi_0^2 - \frac{m^4
\varphi_0^2}{2 m_{p}^2 k^2} + 2 \mu \varphi_0 \cos m t + \left( \frac32 \lambda_0 \varphi_0^2 -
\frac{m^4 \varphi_0^2}{m_{p}^2 k^2} \right) \cos 2 m t \right] \upsilon = 0 \, .
\eer
There appears a term in (\ref{veq1}) which oscillates with frequency $m$. The resonance
instability in the corresponding lowest-frequency zone centered at frequency $\omega =
m/2$ in this case requires the condition
\ber \label{firstres}
\frac{\mu \varphi_0}{m^2} > \frac34 + \frac{k^2}{m^2} - \frac{m^2 \varphi_0^2}{2 m_{p}^2 k^2}
\, ,
\eer
which is incompatible with our assumptions $\mu \varphi_0 \ll m^2$ and $H^2 = m^2
\varphi_0^2/3 m_{p}^2 \ll k^2 \ll m^2$.  However, in the next instability zone centered at
frequency $\omega = m$, there arises a second-order resonance caused by the term $2 \mu
\varphi_0 \cos m t$ in (\ref{veq1}).  Combined with the first-order resonance for the
quartic term, the resulting resonance condition reads
\ber \label{secondres}
k^4 + \left[ \frac94 \lambda_0 \varphi_0^2 - \frac53 \left( \frac{\mu \varphi_0}{m}
\right)^2 \right] k^2 \lesssim \frac{m^4 \varphi_0^2}{m_{p}^2} \, ,
\eer
and represents a generalization of (\ref{quartic}).  The details of derivation of this
expression are presented in the next section.

\section{Parametric resonance to second order}

In this section, we provide an analysis of the parametric resonance in (\ref{veq1}) to
second order of perturbation theory and, in particular, justify the expression
(\ref{secondres}).

Consider a general equation of the form (\ref{veq1}) for a variable $Y$\,:
\ber \label{generic}
\ddot Y + \left[ \omega_0^2 - \epsilon \left( 2 Q_1 \cos \omega t + 2 Q_2 \cos 2 \omega t
\right) \right] Y = 0 \, ,
\eer
where $\omega_0$, $\omega$, $Q_1$ and $Q_2$ are constant in time, and $\epsilon$ is an
explicit small paremeter introduced for considerations of perturbation theory, which
indicates the smallness of $Q_1$ and $Q_2$ relative to $\omega^2$, and which at the end
of calculations can be set to unity. We will be looking for the regions of instability
with respect to the variable proper frequency $\omega_0$ order by order in perturbation
theory in $\epsilon$. It is well known (or can be verified by the theory developed below)
that the resonance zones for equation (\ref{generic}) are located around the frequencies
$\omega_0 = s \omega / 2$ with $s = 1, 2, \ldots$  For this reason, for every natural
$s$, we introduce the quantity $\Delta_s$ defined as
\ber
\epsilon \Delta_s = \left( \frac{s \omega}{2} \right)^2 -\omega_0^2 \, .
\eer
Equation (\ref{generic}) then becomes
\ber \label{gen}
\ddot Y + \left( \frac{s \omega}{2} \right)^2 Y = \epsilon \left( 2 Q_1 \cos \omega t + 2
Q_2 \cos 2 \omega t + \Delta_s \right) Y \, .
\eer

The general asymptotic method developed in \cite{Bogoliubov} consists in looking for
solutions to (\ref{gen}) in the form of expansion in powers of $\epsilon$\,:
\ber \label{Y}
Y = a \cos \frac{s \omega t}{2} + b \sin \frac{s \omega t}{2} + \epsilon u^{(1)} +
\epsilon^2 u^{(2)} + \ldots \, ,
\eer
where $a$ and $b$ are slowly varying functions of time, obeying the system of equations
\ber \label{ab}
\begin{array}{l}
\dot a = \epsilon A^{(1)} (a, b) + \epsilon^2 A^{(2)} (a, b) + \ldots \, , \\
\dot b = \epsilon B^{(1)} (a, b) + \epsilon^2 B^{(2)} (a, b) + \ldots \, ,
\end{array}
\eer
and $u^{(k)}$, $k = 1, 2, \ldots$, are functions of $a$, $b$ and $t$, periodic in the
last argument with period $4 \pi/\omega$\,:
\ber
u^{(k)} \left( a, b, t \right) = \sum_{n \in \mathds{Z}} u^{(k)}_n (a, b) e^{\im n \omega
t/2} \, .
\eer
The method consists in looking for the functions $A^{(k)} (a, b)$, $B^{(k)} (a, b)$ and
$u^{(k)}_n (a, b)$.  Equations (\ref{ab}) then will indicate the domains of instability
with respect to the growth of amplitudes $a$ and $b$.

For our purposes, it will be necessary to develop the theory to second order in
$\epsilon$.  Differentiating (\ref{Y}) and using (\ref{ab}), we find
\ber
\dot Y &=& \frac{s \omega}{2} \left( - a \sin \frac{s \omega t}{2} + b \cos \frac{s
\omega t}{2} \right) + \left( \epsilon A^{(1)} + \epsilon^2 A^{(2)} \right) \cos \frac{s
\omega t}{2} + \left( \epsilon B^{(1)} + \epsilon^2 B^{(2)} \right) \sin \frac{s \omega
t}{2} \nonumber \\ && {} + \epsilon \dot u^{(1)}  + \epsilon^2 \dot u^{(2)} + \epsilon^2
\left( u^{(1)}_a A^{(1)} + u^{(1)}_b B^{(1)} \right) + o \left( \epsilon^2 \right) \, ,
\\
\ddot Y &=& {} - \left( \frac{s \omega}{2} \right)^2 \left(a \cos \frac{s \omega t}{2} +
b \sin \frac{s \omega t}{2} \right) \nonumber \\ && {} - \left( \epsilon A^{(1)} +
\epsilon^2 A^{(2)} \right) s \omega \sin \frac{s \omega t}{2} + \left( \epsilon B^{(1)} +
\epsilon^2 B^{(2)} \right) s \omega \cos \frac{s \omega t}{2} \nonumber \\
&& {} + \epsilon^2 \left( A^{(1)}_a A^{(1)} + A^{(1)}_b B^{(1)} \right) \cos \frac{s
\omega t}{2} + \epsilon^2 \left( B^{(1)}_a A^{(1)} + B^{(1)}_b B^{(1)} \right) \sin
\frac{s \omega t}{2} \nonumber \\ && {} + \epsilon \ddot u^{(1)} + \epsilon^2 \ddot
u^{(2)} + 2 \epsilon^2 \left( \dot u^{(1)}_a A^{(1)} + \dot u^{(1)}_b B^{(1)} \right) + o
\left( \epsilon^2 \right) \, , \label{ddotY}
\eer
where the subscripts `$a$' and `$b$' denote partial derivatives with respect to these
variables, and the dots over $u^{(k)}$ denote partial derivatives with respect to its
last (time) argument.

We substitute (\ref{Y}) and (\ref{ddotY}) into (\ref{gen}) and collect terms with
respective order in $\epsilon$.  The zero-order equation is satisfied automatically. In
the first order in $\epsilon$, we obtain
\ber
\ddot u^{(1)} + \left( \frac{s \omega}{2} \right)^2 u^{(1)} &=& \sum_n \left[ \left(
\frac{s \omega}{2} \right)^2 - \left( \frac{n \omega}{2} \right)^2 \right] u^{(1)}_n
e^{\im n \omega t/2} \nonumber \\
&=& \left( 2 Q_1 \cos \omega t + 2 Q_2 \cos 2 \omega t + \Delta_s \right) \left( a \cos
\frac{s \omega
t}{2} + b \sin \frac{s \omega t}{2} \right) \nonumber \\
&& {} + A^{(1)} s \omega \sin \frac{s \omega t}{2} - B^{(1)} s \omega \cos \frac{s \omega
t}{2} \, . \label{order1}
\eer
The terms with $n = \pm s$ on the left-hand side vanish.  Therefore, the corresponding
terms in the Fourier expansion of the right-hand side should vanish as well.  This leads
to the condition of zero $A^{(1)}$ and $B^{(1)}$ unless $s = 1$ or $s = 2$.  Thus, there
are two potential resonance zones in the first order in $\epsilon$.  We get the following
expressions in these zones:
\ber \label{AB1}
A^{(1)} = \frac{Q_s - \Delta_s}{s \omega} b \, , \qquad B^{(1)} = \frac{Q_s + \Delta_s}{s
\omega} a \, , \qquad s = 1, 2 \, .
\eer
In the first order in $\epsilon$, the two resonant terms in (\ref{gen}) with amplitudes
$Q_1$ and $Q_2$ act independently each in its own resonance zone.

We should also determine the nonzero coefficients $u^{(1)}_n$.  Note that the components
with $n = \pm s$ remain undetermined, but they can be set to zero because the oscillatory
terms with these frequencies are already present as the first two terms in (\ref{Y}). For
the nonzero terms, from (\ref{order1}), we get
\ber \label{u1}
u^{(1)}_3 = - \frac{1}{4 \omega^2} \Bigl[ \left( Q_1 + Q_2 \right) a - \im \left(Q_1 -
Q_2 \right) b \Bigr] \, , \qquad u^{(1)}_5 = \frac{\im}{12 \omega^2} Q_2 b \, , \qquad s
= 1 \, ,
\eer
\ber\label{u2}
u^{(1)}_0 = \frac{Q_1}{\omega^2} a \, , \qquad u^{(1)}_4 = - \frac{Q_1}{6 \omega^2} (a -
\im b) \, , \qquad u^{(1)}_6 = - \frac{Q_2}{16 \omega^2} (a - \im b) \, , \qquad s = 2 \,
,
\eer
and $u^{(1)}_{-q} = \bar u^{(1)}_{q}$ due to the reality of $u^{(1)}$.

In the second order in $\epsilon$, we get the equation
\ber
\ddot u^{(2)} + \left( \frac{s \omega}{2} \right)^2 u^{(2)}
&=& \left( 2 Q_1 \cos \omega t + 2 Q_2 \cos 2 \omega t + \Delta_s \right) u^{(1)} \nonumber \\
&& + A^{(2)} s \omega \sin \frac{s \omega t}{2} - B^{(2)} s \omega \cos \frac{s \omega t}{2} \nonumber \\
&& {} - \left( A^{(1)}_a A^{(1)} + A^{(1)}_b B^{(1)} \right) \cos \frac{s \omega t}{2} -
\left( B^{(1)}_a A^{(1)} + B^{(1)}_b B^{(1)} \right) \sin \frac{s \omega t}{2} \nonumber
\\ && {} - 2 \left( \dot u^{(1)}_a A^{(1)} + \dot u^{(1)}_b B^{(1)} \right) \, .
\eer
Expanding both sides in Fourier series, and taking into account (\ref{AB1}) with an
implication $A^{(1)}_a = B^{(1)}_b = 0$, we obtain
\ber
\sum_n \left[ \left( \frac{s \omega}{2} \right)^2 - \left( \frac{n \omega}{2} \right)^2
\right] u^{(2)}_n e^{\im n \omega t/2} &=& \left( 2 Q_1 \cos \omega t + 2 Q_2 \cos 2 \omega t
+ \Delta_s \right) \sum_p u^{(1)}_p e^{\im p \omega t/2} \nonumber \\
&& {} + A^{(2)} s \omega \sin \frac{s \omega t}{2} - B^{(2)} s \omega \cos \frac{s \omega t}{2} \nonumber \\
&& {} - A^{(1)}_b B^{(1)} \cos \frac{s \omega t}{2} - B^{(1)}_a A^{(1)} \sin \frac{s
\omega t}{2} \nonumber \\
&& {} - \im \omega \sum_p \left( u^{(1)}_{p,a} A^{(1)} + u^{(1)}_{p,b} B^{(1)} \right) p
e^{\im p \omega t/2}  \, .
\eer
We are interested only in the lowest two resonance zones with $s = 1$ and $s = 2$. Again,
the terms with $n = \pm s$ on the left-hand side of the last equation vanish.  Therefore,
the corresponding terms in the Fourier expansion of the right-hand side should vanish as
well, which condition determines the sought functions $A^{(2)}$ and $B^{(2)}$. Taking
into account (\ref{AB1})--(\ref{u2}), we obtain
\ber \label{A2-1}
A^{(2)} &=& \frac{b}{\omega^3} \left[ Q_1^2 - \Delta_1^2 + \frac12
\left( Q_1 - Q_2 \right)^2 + \frac16 Q_2^2 \right]  \, , \\
B^{(2)} &=& \frac{a}{\omega^3} \left[ \Delta_1^2 - Q_1^2 - \frac12 \left( Q_1 + Q_2
\right)^2 \right]  \, , \qquad s = 1\, , \label{B2-1} \\ \nonumber \\
A^{(2)} &=& \frac{b}{8 \omega^3} \left[ Q_2^2 - \Delta_2^2 + \frac43 Q_1^2 + \frac12
Q_2^2 \right]  \, , \label{A2-2} \\
B^{(2)} &=& \frac{a}{8 \omega^3} \left[ \Delta_2^2 - Q_2^2 + \frac{20}{3} Q_1^2 - \frac12
Q_2^2  \right]  \, , \qquad s = 2\, . \label{B2-2}
\eer

One can see that, in both resonance zones, up to second order in $\epsilon$, the
evolution equations have the linear form
\ber \label{AB}
\dot a = A b \, , \qquad \dot b = B a \, ,
\eer
so that the instability region is determined by the condition $A B > 0$.  Setting
$\epsilon = 1$, we obtain, for the coefficients $A$ and $B$ in (\ref{AB}),
\ber \label{A-1}
A &=& \frac{Q_1 - \Delta_1}{\omega} + \frac{1}{\omega^3} \left[ Q_1^2 - \Delta_1^2 +
\frac12 \left( Q_1 - Q_2 \right)^2 + \frac16 Q_2^2 \right] \, , \\
B &=& \frac{Q_1 + \Delta_1}{\omega} + \frac{1}{\omega^3} \left[ \Delta_1^2 - Q_1^2 -
\frac12 \left( Q_1 + Q_2 \right)^2 \right]  \, , \qquad s = 1\, , \label{B-1} \\
\nonumber \\
A &=& \frac{Q_2 - \Delta_2}{2 \omega} + \frac{1}{8 \omega^3} \left[ Q_2^2 -
\Delta_2^2 + \frac43 Q_1^2 + \frac12 Q_2^2 \right]  \, , \label{A-2} \\
B &=& \frac{Q_2 + \Delta_2}{2 \omega} + \frac{1}{8 \omega^3} \left[ \Delta_2^2 - Q_2^2 +
\frac{20}{3} Q_1^2 - \frac12 Q_2^2  \right] \, , \qquad s = 2\, . \label{B-2}
\eer

The boundary of the instability region $AB > 0$ in each of the two zones is determined by
the equations $A = 0$ and $B = 0$.  We can take into account the fact of smallness of the
quantities $Q_s/\omega^2$.  In this case, up to second order in these quantities, we have
$\Delta_s^2 - Q_s^2 = 0$.  Therefore, the roots of (\ref{A-1})--(\ref{B-2}) with respect
to $\Delta_s$ to second order in $Q_s/\omega^2$ can be given by dropping the terms
$\Delta_s^2 - Q_s^2$ from these equations.  For the two zones of interest, we then have
the expressions for the upper ($\Delta_s^+$) and lower ($\Delta_s^-$) boundary values of
the instability zones, depending on the signs of $Q_1$ and $Q_2$:
\ber \label{D1}
\Delta_1^\pm &=& \left\{ \begin{array}{l}
Q_1 + \dfrac{1}{2 \omega^2} \left[ \left( Q_1 - Q_2 \right)^2 + \dfrac13 Q_2^2 \right] \, , \smallskip \\
- Q_1 + \dfrac{1}{2 \omega^2} \left( Q_1 + Q_2 \right)^2 \, ,
\end{array} \right. \\ \nonumber \\
\Delta_2^\pm &=& \left\{ \begin{array}{l}
Q_2 + \dfrac{1}{\omega^2} \left( \dfrac13 Q_1^2 + \dfrac18 Q_2^2 \right)  \, ,  \smallskip \\
- Q_2 - \dfrac{1}{\omega^2} \left( \dfrac{5}{3} Q_1^2 - \dfrac18 Q_2^2 \right) \, .
\end{array} \right.
\label{D2}
\eer

Let us apply these general results to the theory described by equation (\ref{veq1}).  We
have, in this case,
\ber \label{omegas}
\omega = m \, , \qquad \omega_0^2 = m^2 +  k^2 + \frac32 \lambda_0 \varphi_0^2 - \frac{m^4
\varphi_0^2}{2 m_{p}^2 k^2} \, ,
\eer
\ber
Q_1 = {} - \mu \varphi_0 \, , \qquad Q_2 = \left( \frac{m^4 \varphi_0^2}{2 m_{p}^2 k^2} -
\frac34 \lambda_0 \varphi_0^2 \right) \, ,
\eer
so that
\ber
\Delta_1 &=& {} - \frac34 m^2 - k^2 - \frac32 \lambda_0 \varphi_0^2 + \frac{m^4
\varphi_0^2}{2 m_{p}^2 k^2} \, , \\
\Delta_2 &=& {} - k^2 - \frac32 \lambda_0 \varphi_0^2 + \frac{m^4 \varphi_0^2}{2 m_{p}^2
k^2} \, . \label{Deltas}
\eer

In principle, our theory allows for the situation where $Q_1 \gg Q_2$.  In this case, the
resonance condition in the first zone would be dominated by the first-order contribution
in $Q_1$, leading to the condition
\ber
\Delta_1^2 < Q_1^2 \, .
\eer
With the above equations (\ref{omegas})--(\ref{Deltas}), this translates into
(\ref{firstres}).

In the second resonance zone, under the same conditions, we have
\ber
\Delta_2^\pm \approx \left\{ \begin{array}{l}
Q_2 + \dfrac{1}{3 \omega^2} Q_1^2 \, , \smallskip  \\
- Q_2 - \dfrac{5}{3 \omega^2} Q_1^2 \, .
\end{array} \right.
\eer
For our parameters (\ref{omegas})--(\ref{Deltas}), the inequality $\Delta_2^- < \Delta_2
< \Delta_2^+$ implies
\ber
\frac32 \lambda_0 \varphi_0^2 < k^2 + \frac94 \lambda_0 \varphi_0^2 + \frac13 \left(\frac{\mu
\varphi_0}{m} \right)^2 < 2 \left(\frac{\mu \varphi_0}{m} \right)^2 + \frac{m^2
\varphi_0^2}{m_p^2 k^2} \, .
\eer
The lower inequality is satisfied automatically, while the upper one leads to
(\ref{secondres}).

\section{Relationship between the mass of the scalar field and its initial value}
\label{app:mass}

The following conditions need to be simultaneously satisfied in order for $\vphi$ to oscillate
and play the role of dark matter: (i) $\vphi \ll m_p$, (ii) $m \geq H$.

%For the canonical quadratic potential $V(\vphi)=\frac{1}{2}m^2\phi^2$, oscillation frequency is given by $m$. Hence the epoch of oscillation is estimated to be the value of scale factor $a_{\rm osc}$ where $H(a_{\rm osc})\simeq m$. We know that the scalar field starts oscillating in the radiation dominated universe and hence
Since the scalar field starts oscillating during the radiation dominated epoch one finds
\ber
m\simeq H\propto a^{-2}
\Rightarrow a/a_* = \sqrt{(m_*/m)}
\eer
$a_*$ being the epoch marking the onset of oscillations for a given mass $m_*$.  After the scalar
field commences oscillating its average equation of state becomes
$\langle \omega_{\vphi}\rangle=0$ and its density falls off as $\rho_\vphi \propto a^{-3}$.
Consequently
\ber
\frac{\rho}{\rho_*}=\left(\frac{a}{a_*}\right)^{-3}=\left(\frac{m}{m_*}\right)^{{3}/{2}}~.
\label{eq:DM_ratio}
\eer
The initial value of the density is related to the initial scalar field value
by $\rho_i=\frac{1}{2}m^2\vphi_{i}^2$. Substitution in (\ref{eq:DM_ratio}) leads to
\ber
\vphi_i=\vphi_{i*}\left(\frac{m}{m_*}\right)^{-1/4}~.
\label{eq:DMquadPM}
\eer
Finally, our numerical analysis informs us that $\vphi_*=\frac{\sqrt{2\rho_A}}{m}=0.06~m_p$ for
$m_*=10^{-22}$~eV. This initial value of $\vphi_i$ results in  $\Omega_{0m}=0.3$.
This allows us to rewrite (\ref{eq:DMquadPM}) as follows
\ber
\vphi_i=0.06\times \left(\frac{m}{10^{-22}\,{\rm eV}}\right)^{- 1/4}~m_p~,
\label{eq:DMquadPM1}
\eer
which was shown in figure~\ref{fig:DMquadPM}. The corresponding relationship between
$\rho_i$ and $m$ can easily be established using (\ref{eq:DMquadPM1}) and $\rho_i \simeq
\frac{1}{2}m^2\vphi_i^2$.

%%%%%%%%%%%%%%%%%%%%%%%%%%%%%%%%%%%%%%%%%%%%%%%%%%%%%%%%%%%%%%%%%%%%%%%%%%%%%%%%%


\begin{thebibliography}{99}

%\bibitem{Planck-inflation}
%P.~A.~R.~Ade, {\it et.~al.}, \emph{Planck 2013 results. XXII. Constraints on inflation}, arXiv:1303.5082.

\bibitem{linde1}
R. Kallosh and A. Linde, JCAP07 (2013) 002 [arXiv:1306.5220].

\bibitem{linde2}
R. Kallosh, A. Linde and D. Roest, JHEP11, 198 (2013) [arXiv:1311.0472].

\bibitem{sahnicoles_95}
V. Sahni and P. Coles,  Phys.Rept. {\bf 262}, 1-135 (2016) [astro-ph/9505005].

\bibitem{dodelson}
S. Dodelson, {\it Modern Cosmology}, Academic press, (2003).

\bibitem{sahni04}
V. Sahni, {\em Dark matter and dark energy}, Lect. Notes Phys. 653, 141-180 (2004)  [astro-ph/0403324].

\bibitem{DE}
V. Sahni and A.A. Starobinsky, Int. J. Mod. Phys. {\bf D9} 373 (2000);
P.~J.~E. Peebles and B. Ratra, Rev. Mod. Phys. {\bf 75} 559 (2003);
T. Padmanabhan, Phys. Rep. {\bf 380} 235 (2003);
V. Sahni, [astro-ph/0202076], [astro-ph/0502032];
V. Sahni and A.A. Starobinsky, Int. J. Mod. Phys. {\bf D15} 2105 (2006);
E. J. Copeland, M. Sami and S. Tsujikawa, Int. J. Mod. Phys. {\bf D15} 1753 (2006);
R. Bousso, Gen. Relativ. Gravit. {\bf 40}, 607 (2008).

\bibitem{star}
A.A. Starobinsky, \plb {\bf 91}, 99 (1980).

\bibitem{sahni_wang}
V. Sahni and L. Wang, Phys. Rev. D {\bf 62}, 103517 (2000) [astro-ph/9910097].

\bibitem{turner83}
M.S. Turner, Phys. Rev. D {\bf 28}, 1243 (1983).

\bibitem{PQ}
R. Peccei and H.R. Quinn \prl {\bf 38}, 1440 (1977).

\bibitem{zeldovich}
M. Yu. Khlopov, B.A. Malomed and Ya.B. Zeldovich, \mn {\bf 215}, 575 (1985).

\bibitem{DM_early}
S.-J. Sin, \prd {\bf 50}, 3650 (1994);
S.U. Ji and S.-J. Sin, \prd {\bf 50}, 3655 (1994);
J.-W. Lee and I.-G. Koh, \prd {\bf 53}, 2236 (1996).

\bibitem{peebles99}
P.~J.~E. Peebles and A. Vilenkin, \prd {\bf 60}, 103506 (1999);
P.~J.~E. Peebles, Astrophys. J. {\bf 534}, L127 (2000) [astro-ph/0002495].

\bibitem{stein99}
I. Zlatev and P.J. Steinhardt, \plb {\bf 459}, 570 (1999).

\bibitem{hu00}
W.Hu, R. Barkana and A. Gruzinov, \prl {\bf 85}, 1158 (2000).

\bibitem{goodman}
J. Goodman, New Astronomy 5, 103 (2000).

\bibitem{matos00}
T. Matos, F.S. Guzman and D. Nunes, \prd {\bf 62}, 061301 (2000);
T. Matos and L. A. Ure\~na-L\'opez, \prd {\bf 63}, 063506 (2001);
{\em A brief Review of the Scalar Field Dark Matter model},
J. Maga\~na, T. Matos, V. Robles, A. Su\'arez,
[arXiv:1201.6107].

\bibitem{Witten}
L. Hui, J. P. Ostriker, S. Tremaine and E. Witten, [arXiv:1610.08297].

\bibitem{warm}
F. Villaescusa and N. Dalal, JCAP 1103 (2011) 024  [arXiv:1010.3008];
M. Viel, G.D. Becker, J.S. Bolton and M.G. Haehnelt, \prd {\bf 88}, 043502 (2013) [arXiv:1306.2314].

\bibitem{sahni_sen16}
V. Sahni and A.A. Sen,
Eur.Phys.J. {\bf C77}, 225 (2017) [arXiv:1510.09010].
%{\em A new recipe for ΛCDM},  [arXiv:1510.09010].

\bibitem{pulsar}
A. Khmelnitsky and V. Rubakov, JCAP 2 (2014) 019;
N.K. Porayko and K.A. Postnov, \prd {\bf 90}, 062008 (2014) [arXiv:1408.4670].

\bibitem{GW}
A. Aoki and J. Soda, [arXiv:1608.05933].

\bibitem{fuzzy}
A. Arbey, J. Lesgourgues, P. Salati,
Phys.Rev. D {\bf 68}, 023511 (2003)
[astro-ph/0301533];	
F. Siddhartha Guzman, L. A. Urena-Lopez,
Phys.Rev. D {\bf 68}, 024023 (2003)
[astro-ph/0303440];
L. Amendola and B. Barbieri, \plb {\bf 642}, 192 (2006) [hep-ph/0509257];	
A. Bernal and F. Siddhartha Guzman,
\prd {\bf 74}, 103002 (2006)
[astro-ph/0610682];
Jae-Weon Lee,
J.Korean Phys.Soc. 54 (2009) 2622
[arXiv:0801.1442];
L. A. Urena-Lopez,
JCAP 0901 (2009) 014
[arXiv:0806.3093];
T. Matos, J.R Luevano, I. Quiros, L. A. Ure\~na-L\'opez, and J.A. Vazquez,
\prd {\bf 80}, 123521 (2009) [arXiv:0906.0396];
T. Matos, A. Vazquez-Gonzales and J. Magana, \mn {\bf 393},  1359 (2009) [arXiv:0806.0683];
D.J.E. Marsh and P.G. Ferreira, \prd {\bf 82}, 103528 (2010) [arXiv:1009.3501];	
A. P. Lundgren, M. Bondarescu, R. Bondarescu, J. Balakrishna,
Astrophys.J. 715, L35 (2010)
[arXiv:1001.0051];
T. Harko, \prd {\bf 83}, 123515 (2011) [arXiv:1105.5189];
A. Cruz-Osorio, F. Siddhartha Guzman, F. D. Lora-Clavijo,
JCAP 1106 (2011) 029 [arXiv:1008.0027];
V. Lora, J. Magana, A. Bernal, F.J. Sanchez-Salcedo, E.K. Grebel,
JCAP 1202 (2012) 011 [arXiv:1110.2684];
A. J. Christopherson, J. Carlos Hidalgo, K. A. Malik,
JCAP 1301 (2013) 002 [arXiv:1207.1870];
E.J.M. Madarassy, V.T. Toth,
Comput.Phys.Commun. 184, 1339-1343 (2013) , Comput.Phys.Commun. 184, 1339-1343 (2013)  [arXiv:1207.5249];
A. Suárez, V.H. Robles, T. Matos, {\em A Review on the Scalar Field/Bose-Einstein Condensate Dark Matter Model}, [arXiv:1302.0903];
E. Castellanos, C. Escamilla-Rivera, A. Macías, D. Núñez,
JCAP 1411 (2014) no.11, 034 [arXiv:1310.3319];
L. Arturo Ure\~na-L\'opez, \prd {\bf 90}, 027306 (2014) [arXiv:1310.8601];	
T. Harko, \prd {\bf 89}, 084040 (2014) [arXiv:1403.3358];
A. Diez-Tejedor, A. X. Gonzalez-Morales, S. Profumo,
\prd {\bf 90}, 043517 (2014) [arXiv:1404.1054];
L. Visinelli, JCAP 1607 (2016)009 [arXiv:1509.05871];
J.A.R. Cembranos, A.L. Maroto, S. J. Núñez Jareño,
JHEP 1603, 013  (2016) [arXiv:1509.08819];
Jae-Weon Lee, \plb {\bf 756}, 166 (2016) [arXiv:1511.06611];
A. X. Gonzalez-Morales, D.J.E. Marsh, J. Pe\~narrubia, L. A. Ure\~na-L\'opez, [arXiv:1609.05856];
L. Arturo Ure\~na-L\'opez and Alma X. Gonzalez-Morales, JCAP07(2016)048 [arXiv:1511.08195].

\bibitem{marsh}
D. J. E. Marsh, Phys.Rept. {\bf 643}, 1-79 (2016)
[arXiv:1510.07633].

\bibitem{liddle00}
A. R. Liddle and A. Mazumdar,
\prd {\bf 61}, 123507 (2000)
[astro-ph/9912349].

\bibitem{fj97}
P.G. Ferreira and M. Joyce, \prl {\bf 79}, 4740 (1997).

\bibitem{ratra88}
B. Ratra, and P.J.E. Peebles, \prd {\bf 37}, 3406 (1998).

\bibitem{wang}
I. Zlatev, L. Wang and P.J. Steinhardt, \prl {\bf 82} 896 (1999);
P.J. Steinhardt, L. Wang, and I. Zlatev, Phys.Rev. D {\bf 59} 123504 (1999).

\bibitem{Branden}
%M.S. Turner, F. Wilczek and A. Zee, \plb {\bf 125}, 35 (1983);
R. H. Brandenberger, \prd {\bf 32}, 501 (1985).


\bibitem{kamion}
M. C. Johnson and M. Kamionkowski,
\prd {\bf 78}, 063010 (2008)
[arXiv:0805.1748].


\bibitem{monodromy}
E. Silverstein and A. Westphal,
\prd {\bf 78}, 106003 (2008)
[arXiv:0803.3085].

\bibitem{linder}
E. V. Linder, \prd {\bf 91}, no. 12, 123012 (2015) [arXiv:1505.00815].

\bibitem{Kostas_owen}
K. Dimopoulos and C. Owen, [arXiv:1703.00305].


\bibitem{sami2002}
V. Sahni, M. Sami and T. Souradeep, \prd {\bf 65}, 023518 (2002) [gr-qc/0105121].

\bibitem{nojiri2006}
S. Nojiri and S. D. Odintsov,
Phys.Lett. B {\bf 637}, 139 (2006)
[hep-th/0603062].

\bibitem{dutta_scherrer}
S. Dutta and R. J. Scherrer,
\prd {\bf 78}, 083512 (2008)
[arXiv:0805.0763].




\bibitem{preheating}
L.~Kofman, A.~D.~Linde and A.~A.~Starobinsky,
%``Reheating after inflation,''
Phys.\ Rev.\ Lett.\  {\bf 73}, 3195 (1994)
%doi:10.1103/PhysRevLett.73.3195
[hep-th/9405187]; \ Y.~Shtanov, J.~H.~Traschen and R.~H.~Brandenberger,
%``Universe reheating after inflation,''
Phys.\ Rev.\ D {\bf 51}, 5438 (1995)
%doi:10.1103/PhysRevD.51.5438
[hep-ph/9407247].

\bibitem{preheating1}
M. Bastero-Gil, R. Cerezo and Jo\~{a}o G. Rosa, Phys.\ Rev.\ D {\bf 93}, 103531 (2016)
[arXiv:1501.05539].

\bibitem{Bogoliubov}
N.~N.~Bogoliubov and Y.~A.~Mitropolsky, {\it Asymptotic Methods in the Theory of
Non-Linear Oscillations\/}, Gordon and Breach, New York (1961), 537~p.



%\bibitem{log84}
%V. Sahni, PhD thesis, Moscow State University, Moscow, 1984;
%A. Starobinsky and V. Sahni, in {\em Modern Theoretical and Experimental Problems
%of General Relativity}, MGPI press Moscow 1984, p. 77;
%J.R. Primack and G.R. Blumenthal, in {\em Clusters and Groups of Galaxies},
%eds. F. Mardirossian \etal, (Reidel, Drodrecht 1984), p. 435;
%V. Sahni and P. Coles, Phys. Rep. {\bf 262}, 1 (1995).
%
%\bibitem{stagspansion}
%P. Meszaros, \asta {\bf 37}, 225 (1974).

\end{thebibliography}
\end{document}